\documentclass[10pt]{article}

\usepackage{graphicx}
\usepackage{amsmath, amssymb}
\usepackage{geometry}
\usepackage{hyperref}
\usepackage{natbib}
\usepackage{setspace}
\usepackage{caption}
\usepackage[T1]{fontenc}
\usepackage[utf8]{inputenc}
\usepackage{newtxtext,newtxmath}

\usepackage{comment}
\usepackage{tabularx}
\usepackage{dirtytalk}
\usepackage{pdflscape}
\usepackage{caption}
\usepackage{subcaption}
\usepackage{graphicx}
\usepackage{longtable}
\usepackage{lscape}
\usepackage{makecell}
\usepackage{multirow}

\usepackage{float}
\floatstyle{plaintop}
\restylefloat{table}

\title{Ceci N'est Pas un Drone:
Investigating the Impact of Design Representation on Design Decision Making When Using GenAI}

\author{
Zeda Xu$^{1,}$\thanks{Corresponding Author: zedaxu@cmu.edu} \\
Nikolas Martelaro$^1$ \\
Christopher McComb$^1$ \\
\small $^1$Carnegie Mellon University, Pittsburgh, Pennsylvania, USA
}

\date{Dec 04, 2025}

\begin{document}

\maketitle

\begin{abstract}

With generative AI-powered design tools, designers and engineers can efficiently generate large numbers of design ideas. However, efficient exploration of these ideas requires designers to select a smaller group of potential solutions for further development. Therefore, the ability to judge and evaluate designs is critical for the successful use of generative design tools. Different design representation modalities can potentially affect designers' judgments. This work investigates how different design modalities, including visual rendering, numerical performance data, and a combination of both, affect designers' design selections from AI-generated design concepts for Uncrewed Aerial Vehicles. We found that different design modalities do affect designers' choices. Unexpectedly, we found that providing only numerical design performance data can lead to the best ability to select optimal designs. We also found that participants prefer visually conventional designs with axis-symmetry. The findings of this work provide insights into the interaction between human users and generative design systems. 

\end{abstract}


\section{Introduction}

Engineering design is a complex, creative exercise \cite{soria_zurita_survey_2017, song_toward_2020, song_when_2021, xu_adaptation_2024, xu_mind_2025} involving innovative ideation, critical thinking, technical analysis, and iterative problem solving. It also requires designers to make tradeoffs in their search for solutions, balancing between exploration and exploitation in the design space \cite{march_exploration_1991, tabeau_exploration_2017, okamoto_proposal_2022}. Exploration targets original, novel solutions, exemplified by design activities like ideation and brainstorming \cite{march_exploration_1991, gupta_interplay_2006}, whereas exploitation refines existing concepts and prioritizes efficiency, especially during validation and prototyping \cite{march_exploration_1991, gupta_interplay_2006}. Artificial Intelligence (AI) technologies present new opportunities for engineers and designers to achieve improved productivity in both exploration and exploitation \cite{song_when_2021, xu_adaptation_2024}. 

Generative AI can substantially support design exploration by rapidly providing an abundance of feasible design solutions, far beyond what would be possible with human effort alone, further highlighting the importance of the subsequent design selection and exploitation process to materialize those innovations \cite{oh_deep_2019, zhu_generative_2022, zhu_generative_2023, kim_effect_2023}. Engineers and designers, restricted by limited time and resources, must use their expertise and judgment to down-select candidate designs for further validation and prototyping \cite{wallace_methods_1995, gembarski_making_2021}. Therefore, the ability to judge design quality, especially when presented with numerous options, is critical for the successful implementation of generative design tools \cite{hong_generative_2023, chen_how_2025, fang_generative_2025}. 

In the HCI and information science research community, \textit{representation modality} has been used to describe the medium and form for information expression, including visual, auditory, haptic, physical, and other media \cite{hogan_visual_2017, jansen_opportunities_2015, zong_umwelt_2024, bae_making_2022, demir_rethinking_2025}. Drawing from these prior works and considering the engineering context, in the current work, we use \textit{representation modality} to refer to the form in which design information is presented and delivered. Prior studies have shown that the modality in which a design is presented affects people’s perceptions and subjective preferences\cite{reid_impact_2013, detchprohm_effect_2025, barnawal_evaluation_2017, schulze-meesen_impact_2023, derya_ozcelik_buskermolen_effect_2015, padilla_decision_2018}. However, these works mostly focused on graphics information or subjective preferences, rather than on physical engineering designs and objective optimality evaluation. \textbf{The complex nature of engineering design tasks requires both objective information delivery and subjective evaluation of design concept feasibility outside of design data.} Inspired by cognition and neuro-psychology research on judgment and decision-making, we define \textit{judgment ability} in the current work as a designer's capacity to make optimal design decisions and selections, after considering available information and relevant circumstances \cite{rabin_judgment_2007, capucho_judgment_2011, tversky_judgment_1974}. The potential biases introduced by different design representation modalities could weaken designers' judgment abilities and, therefore, reduce the usefulness of generative AI systems in engineering design. 

In this work, we examine generative AI systems from the perspective of what they often afford, particularly in the context of engineering design. Namely, we consider two critical properties of the system in its ability to generate (1) a large set size of the design solutions (GenAI as a design generator), as well as (2) weird or unconventional designs (GenAI for innovative or unhuman-like designs). These attributes characterize the design output from the system, which users directly interact with. 

With or without generative AI systems, it seems an easy conclusion that humans rely on visualization techniques to do design well, given the ubiquity of design visualization in design practices. Studies have shown that design visualization (e.g., sketches) adds information to numerical performance data as a cognitive tool \cite{ullman_importance_1990, goldschmidt_dialectics_1991, suwa_what_1997, larkin_why_1987, stenning_cognitive_1995}, suggesting that designers have heuristics that enable them to extract non-textual information from design visualization. However, AI-generated designs can be "weird" and "unusual", with aesthetics that humans may find unattractive despite performance advantages \cite{loos_towards_2022}. In the context of engineering design, this raises further questions: \textbf{Do engineers actually need design visualization to make good judgments and design choices? Or is the numerical performance data, along with proper data visualization, sufficient? Do design visualization and geometric rendering add value to the design decision-making process beyond numerical performance data?} Research in education has shown that visualization deepens learners' understanding of the subject and improves learning outcomes as an epistemic object \cite{evagorou_role_2015, schoenherr_learning_2024}. It is possible that design visualization and renderings enable human heuristics, allowing engineers and designers to capture features and information that are missed by objective functions or numerical performance data alone. Nevertheless, it is also possible for design visualizations to introduce unwanted bias or design fixation, similar to what prior information visualization and information graphics studies have found, such as exaggerated expectation and illusion of validity \cite{dimara_task-based_2020, tversky_judgment_1975, sharp_performance_1988}. Such decision biases can ultimately lead to sub-optimal design decisions. 

That leads to our central research question: \textbf{Do different design modalities affect human decision-making behavior and their ability to make optimal selections when presented with AI-generated design solutions?} So far, there is insufficient evidence in the design and AI research community to answer this question or to explicitly and systematically examine the impact of design modalities on engineering design decision-making. We hypothesize that (1) different design modalities can affect engineers' decision-making, and (2) using a design visualization (geometric representations of physical products) as a representation modality can limit engineers' ability to identify optimal and novel design solutions. 

To test our hypotheses, we designed and conducted a within-subjects experiment across two studies targeting different populations, examining whether and how different types of design representations, including design visualizations and numerical design performance data, affect engineers’ design choices and their ability to select optimal designs from a list of AI-generated design ideas. 

Our results confirm both of our hypotheses. We found that different design modalities do affect engineers' decision-making when using AI-powered generative design tools. Specifically, in this study, designers provided with only numerical design performance data had the best ability to select optimal designs, outperforming those who saw both the numerical design performance data and the design rendering. We also found that the participants prefer the best-performing designs \textit{as long as} those designs possess traditional and symmetrical appearances. 

This paper contributes empirical evidence on how design representation modalities affect designers' decisions when using AI-powered generative engineering design tools, specifically through:

\begin{itemize}
    \item Empirical findings from two studies with drone hobbyists and STEM students, showing that different representation modalities (visual rendering, numerical data, and visual rendering + numerical data) affect design choices with generative design tools. 

    \item Evidence suggesting that presenting only numerical performance data leads to the most accurate identification of optimal designs, while adding visual renderings can reduce accuracy. Also, a larger number of design choices overwhelms participants, reducing selection accuracy. 
    
    \item Analysis of human heuristics and preferences, revealing that designers prefer conventional, symmetrical designs with good performance. 

    \item Practical implications and recommendations for future development of generative engineering design tools and design comparison tools used with generative AI systems. 
\end{itemize}

As such, the findings of this work could inform the human-AI collaboration loop in engineering design beyond the current GenAI applications, specifically when exploring systems with abundant design options and design novelty.

\section{Related Work}

\subsection{AI in engineering design}

To facilitate solving complex engineering design problems, the engineering design research community has been studying automated design tools since the 1980s \cite{maher_hi-rise_1985, smithers_-based_1989}. With the rapid development of modern AI and Machine Learning (ML) technologies, studies have investigated the implementation of AI and ML in the engineering design process, including design exploration and concept generation \cite{kim_ai_2019, raina_learning_2019, camburn_computer-aided_2020, valdez_framework_2021, zhu_generative_2022, zhu_generative_2023, kim_effect_2023, saadi_generative_2023, khanolkar_mapping_2023, joosten_comparing_2024}, design concept evaluation \cite{camburn_machine_2020, song_decoding_2022, demirel_human-centered_2024}, design optimization \cite{sharpe_comparative_2019, nie_topologygan_2021, behzadi_gantl_2021, senhora_machine_2022, wang_generative_2023, maze_diffusion_2023}, and prototyping and manufacturing \cite{dering_unsupervised_2017, williams_design_2019, qin_research_2022, tercan_machine_2022, kumar_machine_2023}. Those implementations of AI in engineering design are believed to improve the efficiency of the design process \cite{mirhoseini_graph_2021, yuksel_review_2023} and to improve the quality of design solutions \cite{joosten_comparing_2024}. 

More noticeably, recent research has focused on the adoption of AI-powered design tools for design generation \cite{oh_deep_2019, chen_inverse_2021, heyrani_nobari_creativegan_2021, regenwetter_deep_2022}. AI-powered generative design tools can promptly create large sets of design solutions \cite{chen_inverse_2021, heyrani_nobari_creativegan_2021, regenwetter_deep_2022}. With the help of AI generative design tools, engineers and designers can search for potential design solutions more efficiently with a larger scope and at a lower cost \cite{koch_design_2017, oh_deep_2019, camburn_computer-aided_2020, camburn_machine_2020, dering_unsupervised_2017, byrne_application_2025}. In general, generative AI is believed to enhance the design ideation and concept generation process \cite{oh_deep_2019, kim_effect_2023}. However, the resulting abundance of potential design solutions may pose new challenges for designers when selecting optimal designs.

\subsection{Design decision making and judgment ability} 

Concept selection and design decision making are essential to engineering design \cite{wallace_methods_1995, gembarski_making_2021}. Restricted by limited time and resources, engineers and designers constantly face trade-offs and design decision-making in the engineering design process \cite{division_on_engineering_and_physical_sciences_theoretical_2001, otto_trade-off_1991, nickel_contextual_2024}. To make informed decisions, designers must consider various factors and features and seek a balance among them \cite{kalsi_comprehensive_1999, division_on_engineering_and_physical_sciences_theoretical_2001}. As a result, the judgment ability and the ability to evaluate designs are critical for the successful implementation of generative design tools \cite{hong_generative_2023, chen_how_2025, fang_generative_2025}. Experienced engineers and designers can leverage their domain knowledge more effectively to prioritize promising AI suggestions and better utilize this new technology \cite{tambe_reskilling_2025}. It is important to consider the impact on designers' judgment abilities when introducing design tools. 

\subsection{Design representation modalities}

Design representations are frequently used in engineering design ideation, communication, and collaboration \citep{xu_mind_2025, henderson_flexible_1991}. Studies have shown that design representations, especially visual design representations, can serve as boundary objects for effective information exchange within design teams, facilitating collaboration and mitigating misunderstanding \citep{xu_mind_2025, bucciarelli_between_2002, subrahmanian_boundary_2003, kalay_enhancing_2001}. 

Among common design modalities, numerical performance data is an intuitive and straightforward approach to conveying design information accurately and concisely, especially when paired with appropriate data visualization \cite{abi_akle_graphical_2015, araci_trade-off_2017, cibulski_paved_2020}. However, constrained by its textual nature and narrow representation of design features, design information delivered by numerical performance data could be limited in scope (e.g., incomplete or fragmented information on design development and justification \cite{cheng_role_2019, mirabito_feature_2024}). 

In comparison, design visualization (geometric representations of physical products) is another common design modality that is visual-based, more intuitive, and offers a more inclusive representation of physical design features \citep{larkin_why_1987, henderson_flexible_1991, ivanov_visualization_2024}. Visual design representations contain rich information about the design, especially its structure, allowing for accurate and comprehensive interpretation \citep{xu_mind_2025, tsai_how_2017, larkin_why_1987}. As a result, design visualizations are widely used in different stages of the engineering design process \citep{henderson_flexible_1991, haggman_connections_2015, tsai_how_2017, atit_shah_learners_2021, veisz_computer-aided_2012}. It is common practice in the engineering design industry to use visualization of the designs, such as visual renderings or sketches, to aid communications and collaborations \citep{henderson_flexible_1991, haggman_connections_2015, tsai_how_2017, atit_shah_learners_2021}. Research has shown the positive effects of using visualization for engineering design practices, including facilitating ideation, communication, and collaboration, and improving shared understanding \citep{suwa_what_1996, mckoy_influence_2001, tversky_what_2002, tversky_sketches_2003, heiser_sketches_2004, macomber_role_2011, worinkeng_can_2013, xu_mind_2025}. 

However, research has also cautioned about the potential risks of overreliance on design visualizations, including design fixation and detrimental effects on design creativity \cite{jansson_design_1991, atilola_representing_2015, atilola_effects_2016, amann_fixation_1988, viswanathan_design_2013}. These effects can affect designers' judgment, and hinder their willingness and ability to explore novel solutions suggested by generative AI systems. 

\subsection{The impact of design representation modalities on design decision making}

Prior studies show that design modalities could influence people’s perceptions of the design \cite{reid_impact_2013, derya_ozcelik_buskermolen_effect_2015, barnawal_evaluation_2017, schulze-meesen_impact_2023, detchprohm_effect_2025}. Reid et al. \cite{reid_impact_2013} found that different visual representations affect customer subjective preferences and objective measurements of the products, but not their judgments on product function attributes. Detchprohm et al. \cite{detchprohm_effect_2025} showed that the visual quality will not affect the perceived functionality of the product. These works evaluated representations based on perceived functionality and preference, rather than on objective, quantifiable optimality. 

Visual representations can also facilitate communication and feedback compared to textual information \cite{barnawal_evaluation_2017, schulze-meesen_impact_2023}. However, findings on the differences between visual representations are mixed. Barnawal et al. \cite{barnawal_evaluation_2017} suggested that 3D design representations did not outperform 2D ones for communicating design concepts, but they improved usability, while Buskermolen et al. \cite{derya_ozcelik_buskermolen_effect_2015} found that the motion (stills vs. animation) and visual quality did not affect concept comprehension, but visual quality can affect the nature and quality of user feedback. 

Moreover, Padilla et al. \cite{padilla_decision_2018} showed that data graphics influence judgments by invoking different cognitive processes. However, their work focused on abstract data graphics and data visualizations, rather than the visualizations of physical engineering designs. Chen et al. \cite{chen_how_2025} explored textual (ChatGPT) and image-based (Midjourney) generative systems in conceptual design, and found that GenAI can support designers in the problem definition and idea generation stages, but not the idea selection and evaluation stage. The GenAI systems examined in their work emphasize early-stage ideation, whereas generative tools built for physical engineering design must also support design selection under physical constraints, engineering feasibility, and objective optimality. 

Collectively, these studies suggest different design modalities can steer how designers perceive and evaluate designs. However, they provide limited evidence on how design modalities affect the designers' objective evaluation of the optimality of physical engineering designs, especially when GenAI produces large sets of unconventional solutions. Our work addresses this gap by providing empirical evidence on the impact of design modality on designers' ability to identify objectively best-performing designs generated by an AI system.

\section{Methodology}

To test our hypotheses, we designed and conducted a within-subjects experiment across 2 populations to examine whether and how different types of design representations, including design visualizations and numerical design performance data, affect engineers’ design choices and their ability to select optimal designs from a list of AI-generated design ideas. 

\subsection{Participants}
We recruited 156 college students from a U.S. research university. Of the 156 recruited participants, 29 participated in study 1 as self-identified drone pilots and drone hobbyists. The other 127 participants in study 2 are from a sophomore-level engineering design course in the Department of Mechanical Engineering. In this work, we specifically selected drone pilots and hobbyists for their experiential knowledge of drone performance and relevant domain knowledge. We anticipated that they might use this knowledge to augment their decisions. We selected engineering students as proxy participants for early-career engineers capable of understanding the design task and the goal of selecting optimal designs. All participants have the technical capabilities and engineering design knowledge to understand and perform the design tasks. It is a representative group to demonstrate the impact of different design modalities on design decision-making. 

Participation was voluntary, and the participants were compensated with a \$10 Amazon gift card. Participants recruited from the design course are also compensated with course credits. Participants were only allowed to participate in one of the two studies. The ethnicity, age, and gender of the participants did not affect the recruitment process. All participants were over the age of 18 when recruited.

\begin{itemize}
    \item Among the 29 drone pilots and drone hobbyists participants in study 1, 11 identified as women, and 18 identified as men. Also, 7 drone hobbyist participants identified as White, 1 identified as Black or African American, 17 identified as Asian, 3 identified as other, and 1 preferred not to disclose their ethnicity. 

    \item Among the 127 STEM student participants in study 2, 62 identified as women, 57 identified as men, 2 identified as non-binary or third gender, 1 identified as other, and 5 preferred not to identify themselves. Also, 55 STEM student participants identified as White, 8 identified as Black or African American, 35 identified as Asian, 1 identified as Native Hawaiian or Pacific Islander, 20 identified as other, and 8 preferred not to disclose their ethnicity. 
\end{itemize}

\subsection{Study Design}
The university ethics review board approves human-subjects research, and they approved this project. Each study used a Qualtrics online survey in which participants individually considered three UAV (uncrewed aerial vehicles) design problems. The design problems are similar in terms of difficulty and scope, but with different design requirements (e.g., low cost versus low maintenance). For example, 
\begin{quote}
    \textit{"The fire department is using UAVs to monitor wildfires. The UAV should have great hover time and carry at least 10 kg (22.0 lbs) of monitoring equipment. The UAV must rise to a designated height at a fast vertical lift speed. Without sacrificing hover time, more carrying capacity is desirable for carrying more equipment for better coverage. The UAV must also fly steadily and sustain cross-wind and other potential environmental hazards for safety reasons."}
\end{quote}
A detailed problem description and design requirements are provided in Appendix A. 

The UAV designs used in the survey are from the AircraftVerse dataset developed by SRI International and Southwest Research Institute \cite{cobb_aircraftverse_2023}. The UAV designs in this dataset are created using a proprietary generative AI system developed by SRI International \cite{cobb_aircraftverse_2023} with real-world, off-the-shelf components with accurate physical properties. All the UAV designs used in this study are simulated for flight performance using a simulator developed for this work, with support from SRI International and Southwest Research Institute. We then simulated and numerically determined the performance of each UAV design when conducting the tasks described in the design problems. 

Each of the three design problems was accompanied by a set of candidate solutions. For each problem, participants were tasked with selecting the best drone design and were also allowed to share the reasoning for their selection. The number of design solution options is 2, 8, and 16 for the three design problems, respectively. A 4-option case is used for training. For each design problem, design solutions were presented in one of three representation modalities: 1) visual design rendering, 2) numerical design performance data with data visualization, or 3) visual design rendering plus numerical performance data and data visualization. Across modality conditions, the same sets of design solutions appeared, but in random order. 

\begin{figure}
    \centering
    \begin{minipage}[c]{0.45\textwidth}  
        \centering
        \begin{subfigure}[b]{\textwidth}
            \centering
            \includegraphics[width=\textwidth]{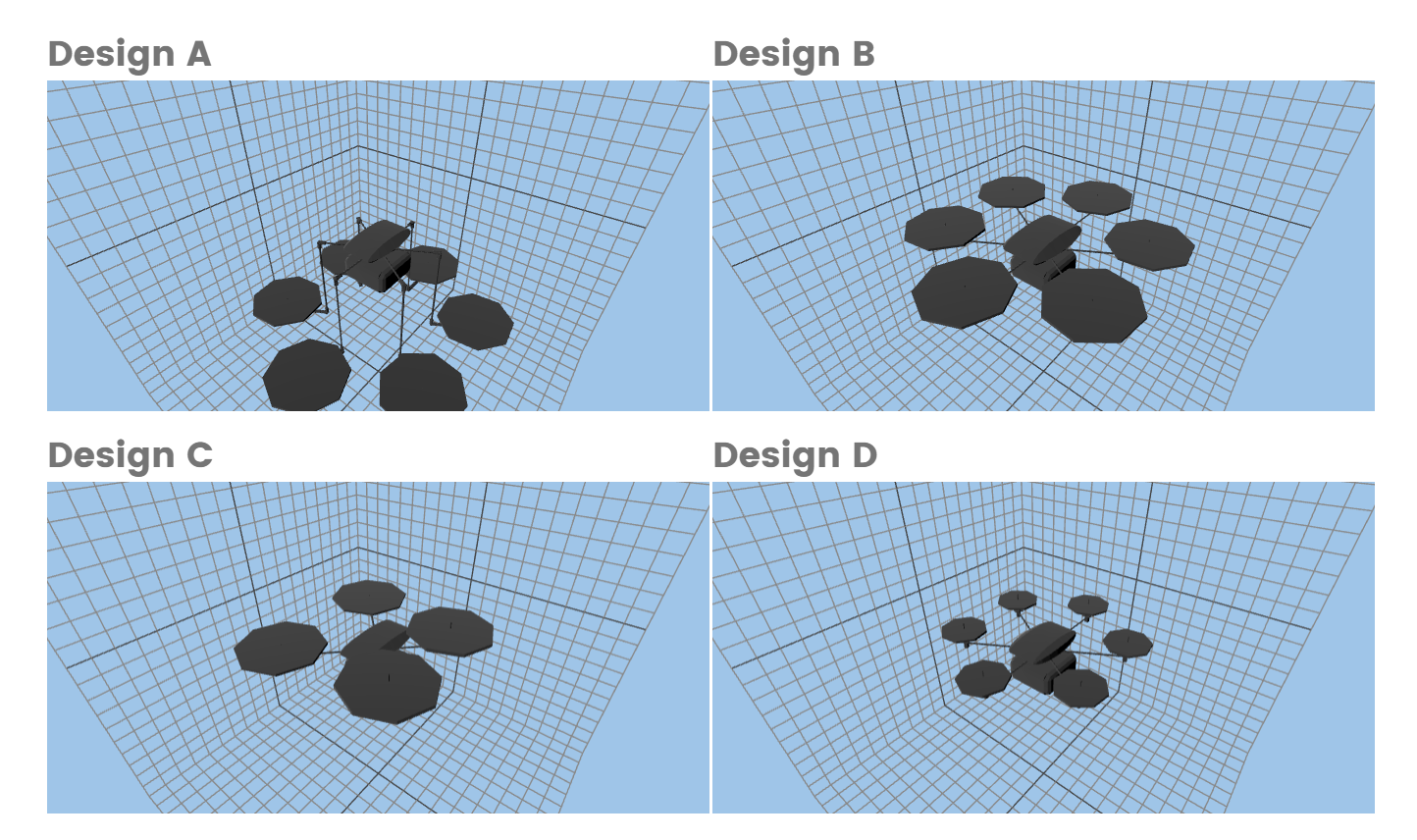}
            \caption{Example design modality - visual rendering. }
            \label{fig:design_modality_example_v}
        \end{subfigure}
        \vspace{1em}
        \begin{subfigure}[b]{\textwidth}
            \centering
            \includegraphics[width=\textwidth]{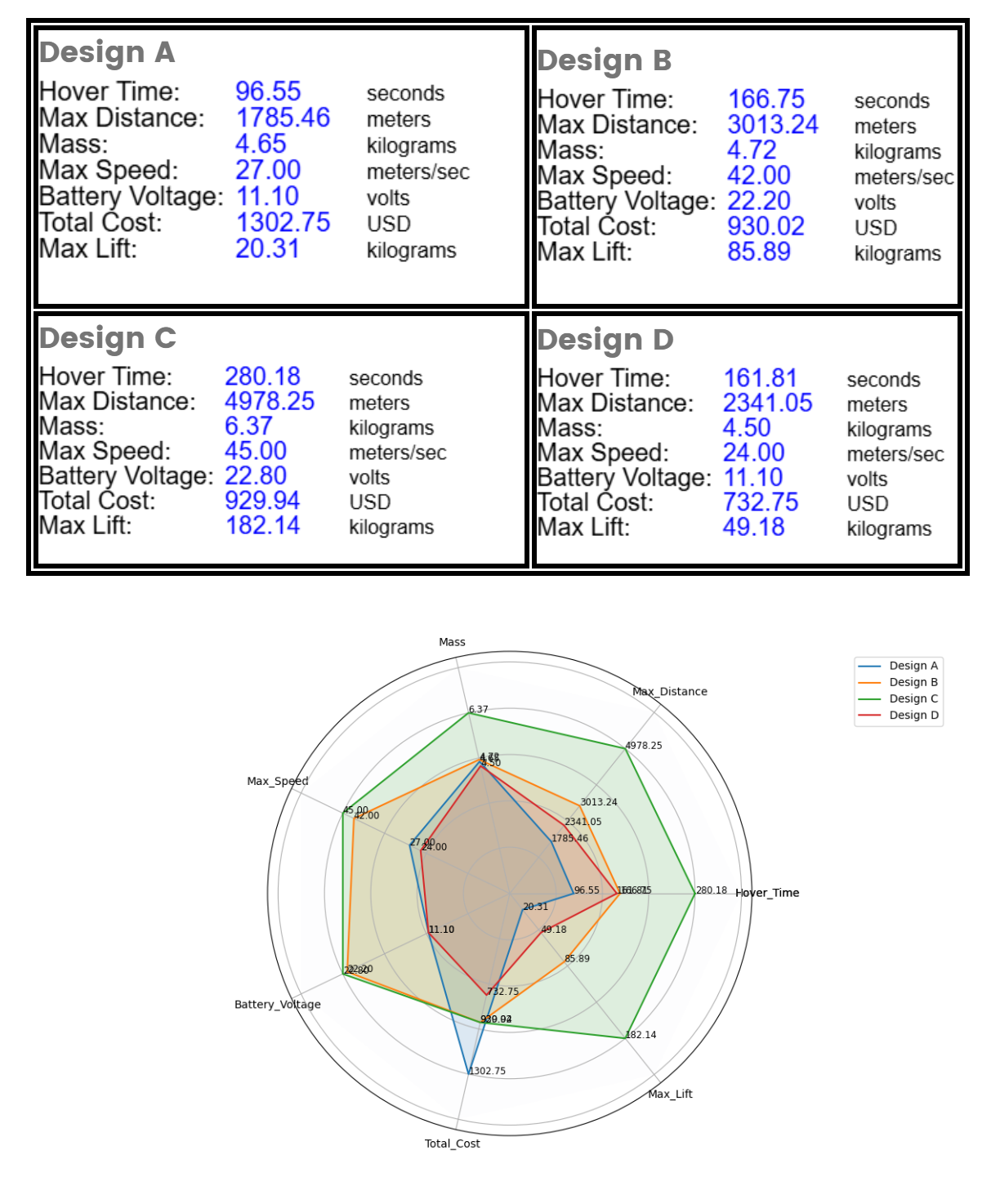}
            \caption{Example design modality - numerical data with data visualization. }
         \label{fig:design_modality_example_d}
        \end{subfigure}
    \end{minipage}
    \hfill
    \begin{minipage}[c]{0.45\textwidth}  
        \centering
        \begin{subfigure}[b]{\textwidth}
            \centering
            \includegraphics[width=\textwidth]{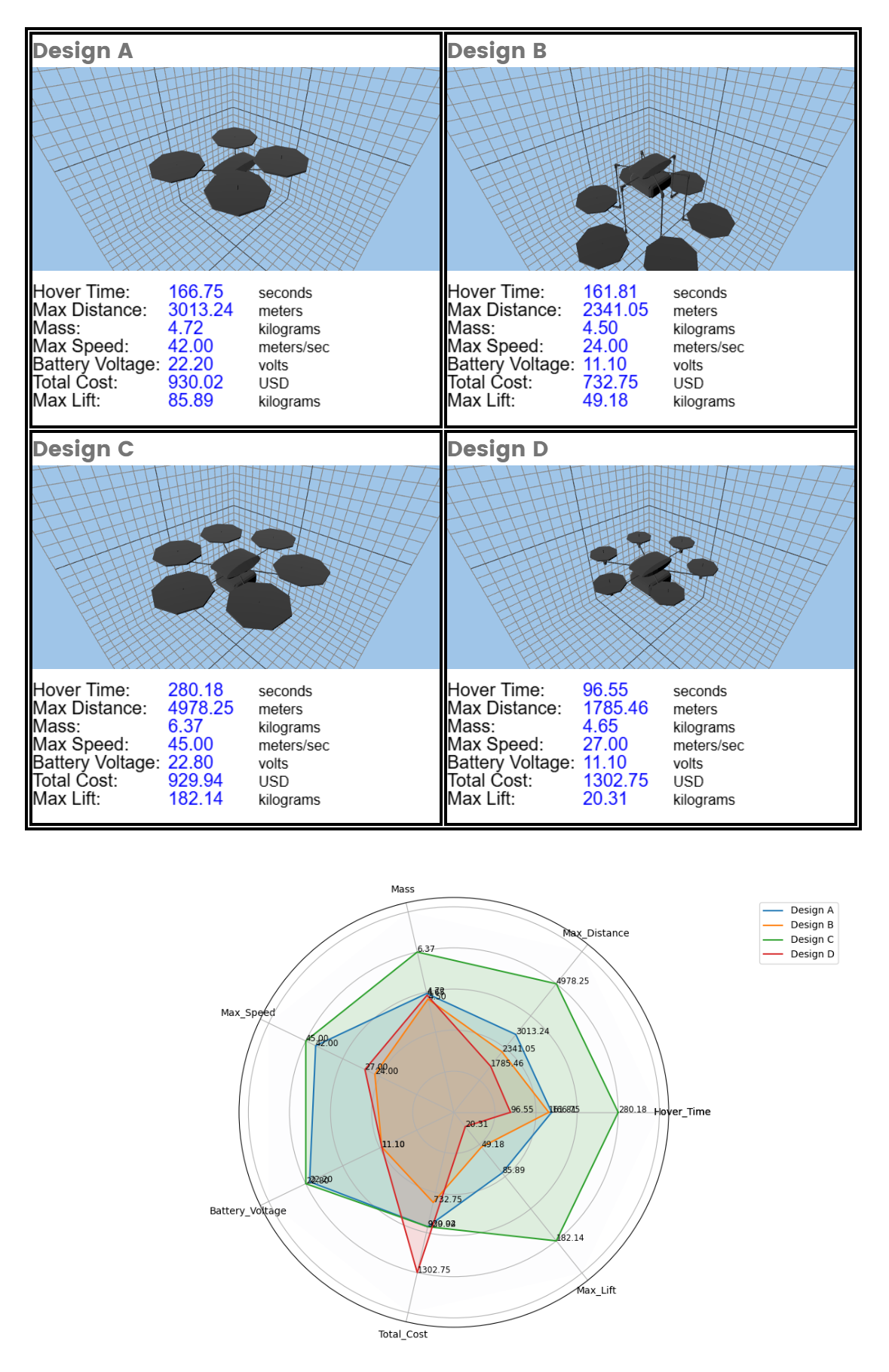}
            \caption{Example design modality - visual rendering with numerical data and data visualization. }
            \label{fig:design_modality_example_m}
        \end{subfigure}
    \end{minipage}

    \caption{Examples of design modalities used in the studies.}
    \label{fig:design_modality_example}
\end{figure}

The visual design rendering is presented with interactive 3D renderings of the UAV designs. For the numerical design performance data with data visualization modality, we presented the performance data as tables along with spider plots. This choice was specifically advised by multiple aerospace professionals who identified the combination of tables and spider plots as a common practice in the aerospace industry. They also recommend against using aligned bar plots. Figure~\ref{fig:design_modality_example} shows an example drone design using these two design modalities, as presented to the participants. The performance data available to the participants are \textit{max hover time}, \textit{max travel distance}, \textit{mass}, \textit{max air speed}, \textit{battery voltage}, \textit{total cost}, and \textit{max lift}. Not all of the listed performance metrics are relevant to every design problem.

The main identifying features and key performance metrics of the drone designs used in this study are shown in Tables~\ref {tab:drone_feature_1}, \ref{tab:drone_feature_2}, and \ref{tab:drone_feature_3} in Appendix B. We characterized each drone by several identifying features: drone designs can be axisymmetric or non-axisymmetric, planar or non-planar, and one-plane or off-plane. Some drones are axisymmetric with respect to the center of the main body, with propellers aligned axisymmetrically on a single horizontal plane. Non-axisymmetric designs lack such axisymmetry. Non-planar means that not all the propellers are on a single horizontal plane. Off-plane designs have at least one propeller on a different horizontal plane than the drone's main body. We define axisymmetric and one-plane designs as conventional designs, and designs with any atypical features (i.e., non-axisymmetric, non-planar, or off-plane) as unusual designs. 

Key performance metrics include \textit{Maximum Lift}, \textit{Effective Lift}, \textit{Hover Time}, and \textit{Total Cost}. \textit{Maximum Lift} is the maximum lift force ($kg$) that the drone can provide. It is one of the most important features in evaluating the performance of the drone designs under the design requirements of the design problems, as it not only affects the amount of weight the drone can lift but also indicates how quickly the drone can move vertically. \textit{Effective Lift} is the maximum weight ($kg$) that the drone can carry, determined by the maximum lift minus the drone's own weight. \textit{Hover Time} describes the maximum time that the drone can stay in the air and maintain the target position with minimal deviation. \textit{Total Cost} is the total cost of the drone. 

The participant groups and design question orders of the two studies are shown in Table~\ref{tab:study_design} below. Each study is detailed in the following subsections.

\subsubsection{Study 1}
For the first study, we targeted drone hobbyists and drone pilots at a U.S. research university. In total, 29 student drone hobbyists were recruited. In this study, the three design problems were presented in the order of 2-option, 8-option, and 16-option. 

\subsubsection{Study 2}
For the second study, we recruited 127 STEM students at the U.S. research university through an engineering design related course. The second study has two different survey design conditions. The first condition, Study 2A (63 participants), presents the design problems in the order of 2-option, 8-option, and 16-option, identical to the first study. The second version of the survey, Study 2B (64 participants), presents the design problems in the order of 2-option, 16-option, and 8-option. Both versions of the survey are identical in all aspects other than the order in which the design problems are shown. The design problems and the design options are the same as those used in the first study. 
\footnotemark

\footnotetext{In Study 1, we initially observed a drastic difference in participants' responses between the 8-option and the 16-option design problems, with our initial analysis method for design optimality. To determine whether the irregularity is significantly influenced by the order in which the design problems were presented, we conducted the second study with two different design problem orders. However, after data collection, we switched to the current Pareto-TOPSIS method for more objective and accurate optimality evaluation, and the irregularity is no longer observed. }

\begin{table}[H]
    \centering
    \begin{tabularx}{0.8\textwidth} { 
    | >{\centering\arraybackslash}p{0.15\textwidth} 
    | >{\centering\arraybackslash}p{0.25\textwidth} 
    | >{\centering\arraybackslash}X | }
    
    \hline
    \textbf{Study Number} & \textbf{Participant Group} \newline (Number of Participants) & \textbf{Design Question Order} \newline (Number of Design Options for Each Design Problem) \\
    \hline
    Study 1
    & Drone Hobbyists (29) & 2 Options -> 8 Options -> 16 Options \\
    \hline
    Study 2A
    & STEM students (63) & 2 Options -> 8 Options -> 16 Options \\
    \hline
    Study 2B
    & STEM students (64) & 2 Options -> 16 Options -> 8 Options \\
    \hline

\end{tabularx}
\caption{Study Designs}
\label{tab:study_design}
\end{table}

\subsection{Procedure}

Participants completed the study using an online survey tool (Qualtrics). After giving consent, participants began the study and were instructed to consider three design problems about UAVs with different design requirements, with design options shown in different modalities. Participants are informed that the designs are automatically generated by an AI system and that we need their expertise to evaluate the feasibility of these solutions. The participants are also informed that the designs are tested in an advanced simulator, but they still need to utilize their engineering experience and knowledge to evaluate the designs, considering real-world scenarios. The study then asks the participant to choose the most optimal drone design from the provided list of options and briefly explain their choice in a textual response.

\subsection{Measurement and Data Analysis}

We analyzed participants' responses using three quantitative measures: (1) response entropy, (2) response changes across modalities, and (3) response accuracy relative to optimal designs. 

\subsubsection{Multiple Choice Question Response Entropy}

Before performing any analysis or statistical tests on the multiple-choice question responses, we first confirm that the participants are making informed decisions. To test if the question responses from the participants are rooted in the provided design information rather than randomness, we use Shannon entropy as a measurement of the level of information obtained by the participants. A smaller entropy value corresponds to greater order and less randomness in the participants' choices, suggesting the participants are making similar decisions instead of random choices. The entropy is calculated as 

\begin{equation}
    H = - \sum_{i=1}^{n} p_i \ln p_i 
\end{equation}
\\
where H is the Shannon entropy and $p_i$ is the probability of the $i$-th design being chosen. We estimate these probabilities simply as

\begin{equation}
    p_i = \frac{ 
    \text{\# times design } i \text{ was chosen} 
    }{
    \text{total choices made}
    }
\end{equation}

\subsubsection{Multiple Choice Question Response Change Across Modalities}

To understand the impact of different design modalities, we test whether the participants are making different design choices with different design modalities for the same design problem. Specifically, we seek to examine how the design choices changed between modalities. We illustrated and analyzed changes in participants' design choices across design representation modalities, using figures showing the number of participants who changed design choices when modalities changed. A high number of participants who changed their design choices indicates that different design modalities influenced their decisions. 

In addition, we illustrated and analyzed participants’ selection of conventional designs (axisymmetric and one-plane) versus unusual designs (non-axisymmetric, non-planar, or off-plane) with figures showing the number and percentage of participants who picked those designs. A higher number and percentage of participants selecting those designs reveal their preference for either conventional or unusual designs when presented with different design modalities.

\subsubsection{Multiple Choice Question Response Accuracy}

In addition to unveiling whether different design modalities affected participants' design decisions, we examine how different design modalities affect the participants' ability to select better designs.

In order to consistently select the best design, we employ the Pareto-TOPSIS (Technique for Order of Preference by Similarity to Ideal Solution) method \cite{hwang_multiple_1981, hwang_new_1993, hu_new_2023}. Pareto-TOPSIS ranks candidate options based on their weighted distance to the ideal and the worst options, and is widely used in engineering design and HCI research as an objective means of establishing optimality \cite{hu_new_2023, alizadeh_data-driven_2019, souaille_interactive_2022, bertoni_iterative_2019, wang_enhanced_2021, chatterjee_probabilistic_2017}. The option closest to the ideal and farthest from the worst is deemed the most optimal. In this work, we applied the same weight to all the design objectives in the design problems, namely, \textit{Hover Time}, \textit{Maximum Lift}, \textit{Effective Lift}, and \textit{Total Cost}. 

The Pareto-TOPSIS optimal solutions for problems with 2, 8, and 16 options are design 18393, design 20155, and design 20985, respectively, and they are indicated with an asterisk in Tables~\ref{tab:drone_rendering_1}-\ref{tab:drone_feature_3} in Appendix B. We then calculated the accuracy of participants selecting the Pareto-TOPSIS optimal designs and performed ANOVA (Analysis of Variance) tests on the accuracies to unveil differences across experimental conditions. We also performed follow-up t-tests, with a Bonferroni correction adjusted alpha value, to compare participants' accuracy with different design modalities as post hoc tests. Higher accuracy in selecting Pareto-TOPSIS optimal designs indicates a stronger ability to select better designs when participants are provided with that design modality.

\section{Results}

The results of each study will be shown separately in the following sub-sections. Each study subsection includes quantitative analyses of multiple-choice question responses, including response entropy, changes in design choice across modalities, and accuracy. 

\subsection{Study 1 - Drone Hobbyists}
The participants in Study 1 are drone hobbyists. The number of participants that have chosen each design and the participants' choice transitions are shown in Figures \ref{fig:sankey_cmu_flyer_problem_1.png}, \ref{fig:sankey_cmu_flyer_problem_2.png}, and \ref{fig:sankey_cmu_flyer_problem_3.png} in Appendix C. 

\subsubsection{Multiple Choice Question Response Entropy}
The entropy values of the design choices for each design problem are calculated (Design problem with 2 options: $H_{max}=1$, $H_{Visual} = 0.401$, $H_{Data} = 0.251$, $H_{Visual + Data} = 0.678$. 
Design problem with 8 options: $H_{max}=3$, $H_{Visual} = 1.273$, $H_{Data} = 1.539$, $H_{Visual + Data} = 1.500$. 
Design problem with 16 options: $H_{max}=4$, $H_{Visual} = 1.401$, $H_{Data} = 1.497$, $H_{Visual + Data} = 1.551$). Overall, the entropy values are relatively small, indicating low randomness in the participants' answers, suggesting the participants are making informed decisions based on the provided design information.

\subsubsection{Multiple Choice Question Response Change Across Modalities}
\begin{figure}[H]
    \centering
    \includegraphics[width=0.5\linewidth]{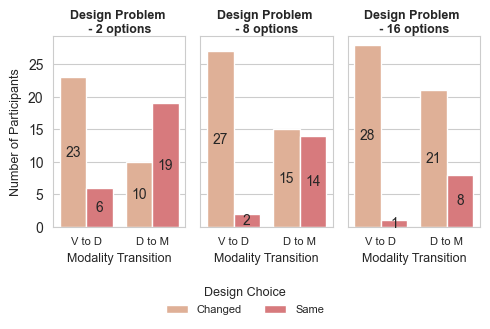}
    \caption{Study 1: Participants' choice of design for consecutive questions in three design problems. V is with visual rendering, D is with numerical performance data, and M is with both visual rendering and numerical performance data. }
    \label{fig:same_vs_different_choices_cmu_flyer_problem}
\end{figure}

The consistency of participants' design choices when presented with different design modalities is shown in Figure~\ref{fig:same_vs_different_choices_cmu_flyer_problem}. Most participants' design choices changed when the design modality changed from visual rendering to numerical performance data. Design modalities did affect participants' decision-making here. However, there are no consistent changes found when the design modality changed from numerical performance data to the mixed modality with both visual rendering and numerical performance data. The transitions of participants' design choice changes are shown in Figures \ref{fig:sankey_cmu_flyer_problem_1.png}, \ref{fig:sankey_cmu_flyer_problem_2.png}, and \ref{fig:sankey_cmu_flyer_problem_3.png}. 

\begin{figure}[H]
    \centering
    \includegraphics[width=0.5\linewidth]{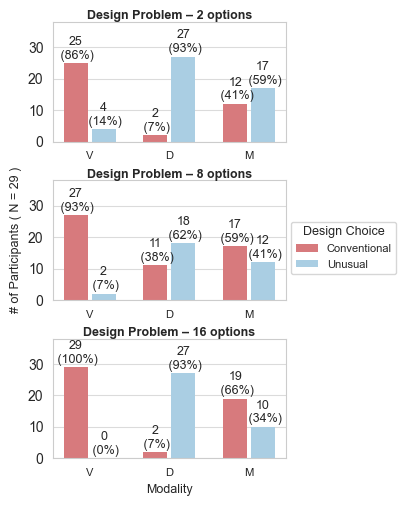}
    \caption{Study 1: Participants' choice for conventional design vs unusual design in percentage. }
    \label{fig:design_choice_cmu_flyer}
\end{figure}

Participants' selection of conventional design (axisymmetric and one-plane) vs unusual design (non-axisymmetric, non-planar, or off-plane) is shown in Figure~\ref{fig:design_choice_cmu_flyer}. The participants showed a strong preference for conventional designs when they were provided with only visual renderings. Also, fewer participants picked the unusual designs when visual renderings became available, when transitioning from numerical performance data to the mixed modality (both visual rendering and numerical performance data). 

\subsubsection{Multiple Choice Question Response Accuracy}
\begin{figure}[H]
    \centering
    \includegraphics[width=0.5\linewidth]{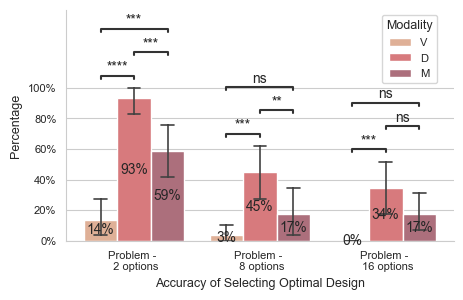}
    \caption{Study 1: Participants' accuracy on selecting the optimal designs. Error bars represent a 95\% confidence interval. 
    p-value annotation legend:
    ns: 1.70e-02 < p <= 1.00e+00; 
    \quad *: 1.00e-02 < p <= 1.70e-02; 
    \quad **: 1.00e-03 < p <= 1.00e-02; 
    \quad ***: 1.00e-04 < p <= 1.00e-03; 
    \quad ****: p <= 1.00e-04. 
    Participants are drone hobbyists. }
    \label{fig:cmu_flyer_acc_res_pareto}
\end{figure}

The average results of drone hobbyist participants' accuracy on selecting the optimal designs are shown in Figure~\ref{fig:cmu_flyer_acc_res_pareto}, with error bars indicating a 95\% confidence interval. 

We conducted ANOVA tests on the accuracy for each design problem. There are significant differences in participants' accuracy in selecting the optimal designs when different design modalities are offered (Design problem with 2 options: $F(2, 84) = 31.207$, $p < 0.001$, $\eta _p^2 = 0.426$. Design problem with 8 options: $F(2, 84) = 8.809$, $p < 0.001$, $\eta _p^2 = 0.173$. Design problem with 16 options: $F(2, 84) = 6.774$, $p = 0.002$, $\eta _p^2 = 0.139$). The results indicate that the design modality does affect participants' abilities to choose the better designs. 

In addition, we performed two-way ANOVA tests on the design modality and the number of design options (design problem) to see if the number of design options provided would change participants' decision-making results (Effect of Design Problem: $F(2, 252) = 26.605, p < 0.001$; Interaction Effect: $F(4, 252) = 3.268, p = 0.012$). Both the design modality and the number of design options affect the accuracy of participants, and the interaction effect between the design modality and the design problem is also significant. 

Further, we performed $t$-tests between experiment conditions, with a Bonferroni correction adjusted alpha value of 0.017, and the results are shown in Table \ref{tab:cmu_flyer_ttest}. In more than half of the pairwise comparisons, there are significant differences in participants' accuracies for selecting the optimal designs between the visual rendering and numerical data conditions, between the numerical data and mixed modality conditions, and between the visual rendering and mixed modality conditions. However, there are no significant differences found between numerical data and the mixed modality conditions for picking the optimal designs in the design problem with 16 design options. There are also no significant differences found between visual rendering and the mixed modality conditions for picking the optimal designs in both the design problem with 8 design options and the design problem with 16 design options. 

\begin{table}[H]
    \centering
    
    \begin{tabularx}{0.7\textwidth} { 
    | >{\centering\arraybackslash}p{0.15\textwidth} 
    | >{\centering\arraybackslash}p{0.1\textwidth} 
    | >{\centering\arraybackslash}X | }
    
    \hline
    \textbf{Comparison}\newline \textbf{Group} & \textbf{Design} \newline \textbf{Problem} & \textbf{T-test results} \\
    \hline
    \multirow{3}{*}{\makecell{Visual vs \\ Data}} 
    & 2 options & t(28) = 10.360, p < 0.001, Cohen's d = 1.924 \\ 
    \cline{2-3}
    & 8 options & t(28) = 3.923, p < 0.001, Cohen's d = 0.728 \\
    \cline{2-3}
    & 16 options & t(28) = 3.839, p < 0.001, Cohen's d = 0.713 \\
    \hline
    \multirow{3}{*}{\makecell{Data vs \\ Visual + Data}}
    & 2 options & t(28) = 3.839, p < 0.001, Cohen's d = 0.713 \\
    \cline{2-3}
    & 8 options & t(28) = 2.816, p =0.009, Cohen's d = 0.523 \\
    \cline{2-3}
    & 16 options & t(28) = 1.983, p = 0.057, Cohen's d = 0.368 \\
    \hline
    \multirow{3}{*}{\makecell{Visual vs \\ Visual + Data}}
    & 2 options & t(28) = 4.218, p < 0.001, Cohen's d = 0.783 \\
    \cline{2-3}
    & 8 options & t(28) = 1.684, p = 0.103, Cohen's d = 0.313 \\
    \cline{2-3}
    & 16 options & t(28) = 2.415, p = 0.023, Cohen's d = 0.448\\
    \hline

\end{tabularx}
\caption{Study 1: T-test results on the effect of design modalities on the accuracy of participants in selecting the optimal designs. }
\label{tab:cmu_flyer_ttest}
\end{table}

\subsection{Study 2 - Engineering Students}

The participants in Study 2A and Study 2B are engineering students at the aforementioned U.S. research university. The design problems for Study 2A are presented in the order of 2-option, 8-option, and 16-option, and those in Study 2B are presented in the order of 2-option, 16-option, and 8-option. 

The results from Study 2A and Study 2B are first compared based on the accuracy of selecting the optimal designs. We performed two-way ANOVA tests on the design modality and the order in which the design questions are presented to see if the order of design questions would affect participants' choices (Design problem with 2 options: Effect of Question Order $F(1, 375) = 0.172, p = 0.678$, Interaction Effect $F(2, 375) = 0.322, p = 0.725$. Design problem with 8 options: Effect of Question Order $F(1, 375) = 0.011, p = 0.917$, Interaction Effect $F(2, 375) = 0.035, p = 0.965$. Design problem with 16 options: Effect of Question Order $F(1, 375) = 2.235, p = 0.136$, Interaction Effect $F(2, 375) = 1.924, p = 0.147$). The order in which the design questions were presented did not significantly affect the accuracy of participants. 

Further, we performed an Extra‑Sum‑of‑Squares F–test on the participants' accuracy to determine if the order of design questions affects participants' overall responses. We compared two nested regression models, with the reduced model using "modality" and "question list" as predictors, and the full model using an additional predictor, "order of question list." The results suggest that there is no evidence that the additional variable (i.e., "order of question list") adds predictive power ($F(1, 1135) = 0.351, p = 0.553$). The patterns of responses collected from Study 2A and Study 2B are not statistically different. \textbf{Therefore, the results of Study 2A and Study 2B will be combined as one study and presented together.} The number of participants that have chosen each design and the participants' choice transitions are shown in Figures \ref{fig:sankey_cmu_class_comb_problem_1.png}, \ref{fig:sankey_cmu_class_comb_problem_2.png}, and \ref{fig:sankey_cmu_class_comb_problem_3.png} in Appendix C. 

\subsubsection{Multiple Choice Question Response Entropy}
The entropy values of the design choices for each design problem are calculated (Design problem with 2 options: $H_{max}=1$, $H_{Visual} = 0.330$, $H_{Data} = 0.112$, $H_{Visual + Data} = 0.640$. 
Design problem with 8 options: $H_{max}=3$, $H_{Visual} = 1.584$, $H_{Data} = 1.445$, $H_{Visual + Data} = 1.688$. 
Design problem with 16 options: $H_{max}=4$, $H_{Visual} = 1.764$, $H_{Data} = 1.744$, $H_{Visual + Data} = 1.973$). Overall, the entropy values are relatively small, indicating low randomness in the participants' responses, suggesting the participants are making informed decisions based on the provided design information. 

\subsubsection{Multiple Choice Question Response Change Across Modalities}
\begin{figure}[H]
    \centering
    \includegraphics[width=0.5\linewidth]{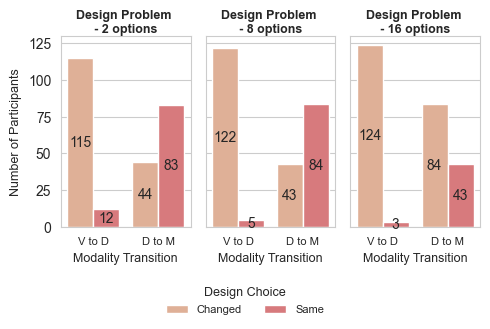}
    \caption{Study 2: Participants' choice of design for consecutive questions in three design problems. V is with visual rendering, D is with numerical performance data, and M is with both visual rendering and numerical performance data. }
    \label{fig:same_vs_different_choices_cmu_class_comb_problem}
\end{figure}

The consistency of participants' design choices when presented with different design modalities is shown in Figure~\ref{fig:same_vs_different_choices_cmu_class_comb_problem}. Most participants' choices changed when the design modality changed from visual rendering to numerical performance data. Thus, design modalities did affect participants' decision-making here. However, changes are inconsistent when the design modality changed from numerical performance data to the mixed modality with both visual rendering and numerical performance data. The changes in participants' design choices are shown in Figures~\ref{fig:sankey_cmu_class_comb_problem_1.png}, \ref{fig:sankey_cmu_class_comb_problem_2.png}, and \ref{fig:sankey_cmu_class_comb_problem_3.png}. 

\begin{figure}[H]
    \centering
    \includegraphics[width=0.5\linewidth]{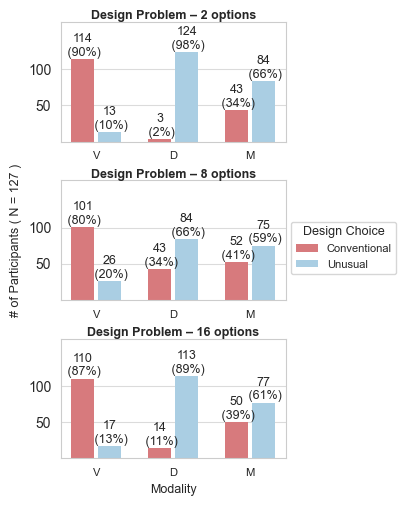}
    \caption{Study 2: Participants' choice for conventional design vs unusual design in percentage. }
    \label{fig:design_choice_cmu_class}
\end{figure}

Participants' selection of conventional designs (axisymmetric and one-plane) versus unusual designs (non-axisymmetric, non-planar, or off-plane) is shown in Figure~\ref{fig:design_choice_cmu_class}. The participants showed a strong preference for conventional designs when they were provided with only visual renderings. Also, fewer participants picked unusual designs when visual renderings became available (i.e., when transitioning from numerical performance data to the mixed modality). 

\subsubsection{Multiple Choice Question Response Accuracy}
\begin{figure}[H]
    \centering
    \includegraphics[width=0.5\linewidth]{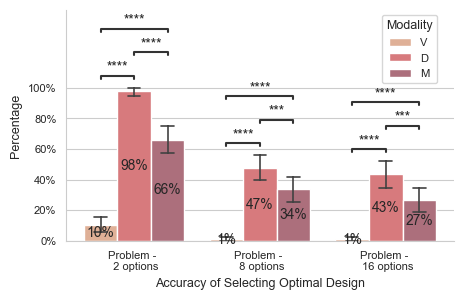}
    \caption{Study 2: Participants' accuracy on selecting the optimal designs. Error bars represent 95\% confidence interval. 
    p-value annotation legend:
    ns: 1.70e-02 < p <= 1.00e+00; 
    \quad *: 1.00e-02 < p <= 1.70e-02; 
    \quad **: 1.00e-03 < p <= 1.00e-02; 
    \quad ***: 1.00e-04 < p <= 1.00e-03; 
    \quad ****: p <= 1.00e-04. 
    Participants are STEM students. }
    \label{fig:cmu_class_comb_acc_res_pareto}
\end{figure}

The average results of engineering student participants' accuracy on selecting the optimal designs are shown in Figure~\ref{fig:cmu_class_comb_acc_res_pareto}, with error bars indicating a 95\% confidence interval.  

We conducted ANOVA tests on the accuracy for each design problem. There are significant differences in participants' accuracy in selecting the optimal designs when different design modalities are offered (Design problem with 2 options: $F(2, 378) = 218.552$, $p < 0.001$, $\eta _p^2 = 0.536$. Design problem with 8 options: $F(2, 378) = 44.940$, $p < 0.001$, $\eta _p^2 = 0.192$. Design problem with 16 options: $F(2, 378) = 38.645$, $p < 0.001$, $\eta _p^2 = 0.170$). The results indicate that the design modality would affect participants' abilities to choose the "better" designs. 

Additionally, we performed two-way ANOVA tests on the design modality and the number of design options (design problem) to see if the number of design options provided would change participants' decision-making results (Effect of Design Problem: $F(2, 1134) = 95.540, p < 0.001$. Interaction Effect: $F(4, 1134) = 14.233, p < 0.001$). Both the design modality and the number of design options affect the accuracy of participants, and the interaction effect between the design modality and the design problem is also significant. 

Further, we performed $t$-tests between experiment conditions, with a Bonferroni correction adjusted alpha value of 0.017, and the results are shown in Table~\ref{tab:cmu_class_comb_ttest}. In general, there are significant differences in participants' accuracies for selecting the optimal designs between the visual rendering and numerical data conditions, between numerical data and mixed modality conditions, and between visual rendering and mixed modality conditions. 

\begin{table}[H]
    \centering
    
    \begin{tabularx}{0.7\textwidth} { 
    | >{\centering\arraybackslash}p{0.15\textwidth} 
    | >{\centering\arraybackslash}p{0.1\textwidth} 
    | >{\centering\arraybackslash}X | }
    
    \hline
    \textbf{Comparison}\newline \textbf{Group} & \textbf{Design} \newline \textbf{Problem} & \textbf{T-test results} \\
    \hline
    \multirow{3}{*}{\makecell{Visual vs \\ Data}} 
    & 2 options & t(126) = 26.071, p < 0.001, Cohen's d = 2.313 \\ 
    \cline{2-3}
    & 8 options & t(126) = 10.140, p < 0.001, Cohen's d = 0.900 \\
    \cline{2-3}
    & 16 options & t(126) = 9.358, p < 0.001, Cohen's d = 0.830 \\
    \hline
    \multirow{3}{*}{\makecell{Data vs \\ Visual + Data}}
    & 2 options & t(126) = 7.110, p < 0.001, Cohen's d = 0.631 \\
    \cline{2-3}
    & 8 options & t(126) = 3.405, p < 0.001, Cohen's d = 0.302 \\
    \cline{2-3}
    & 16 options & t(126) = 3.415, p < 0.001, Cohen's d = 0.303 \\
    \hline
    \multirow{3}{*}{\makecell{Visual vs \\ Visual + Data}}
    & 2 options & t(126) = 11.902, p < 0.001, Cohen's d = 1.056 \\
    \cline{2-3}
    & 8 options & t(126) = 7.624, p < 0.001, Cohen's d = 0.677 \\
    \cline{2-3}
    & 16 options & t(126) = 6.394, p < 0.001, Cohen's d = 0.567 \\
    \hline

\end{tabularx}
\caption{Study 2: T-test results on the effect of design modalities on the accuracy of participants in selecting the optimal designs. }
\label{tab:cmu_class_comb_ttest}
\end{table}

\section{Discussion}

\subsection{Design modalities affect design decision making}

We found that different design modalities appear to affect engineers' decision-making, and that using only visual design renderings as a design modality can limit engineers' ability to identify the optimal design solutions when using AI-powered generative design tools. 

For both participant groups, most participants changed their design choices when the design modality switched from visual rendering to numerical performance data. Also, a significant portion of participants changed their design choices when the design modality switched from numerical performance data to the mixed modality. Since the design options remained the same, these changes indicate that design modalities did affect participants' design choices. Participants changed their minds and possibly selected based on different factors when presented with different design modalities. 

The difference between design modalities is also evident in participants' ability to select optimal designs. The three design modalities showed significantly different levels of benefits for selecting the optimal design. Participants identified optimal designs with much higher accuracy when shown only numerical performance data than when shown visual renderings or mixed modality. This result also demonstrates people's ability to read and interpret design information from numerical data and spider plots, even though spider plots face criticism from the data visualization research community, including inconsistent areas and shapes caused by axis ordering, misleading area size, and deceptive importance of irrelevant options due to dimension normalization \cite{feldman_filled_2013, heijungs_two_2022, duan_origami_2023, abeynayake_efficacy_2023}. 

Participants' accuracy in selecting optimal designs was low when they saw only visual renderings. This might suggest that visual renderings alone either provide limited information about performance or convey distorted cues (e.g., propellers appearing smaller when the UAV body is larger). However, an alternative explanation is that participants gained additional information from the visual renderings that is not captured by the performance data or by our simulator (e.g., serviceability, vibration control, performance in cross-winds). In that case, the difference in accuracy in selecting optimal designs in the visual rendering condition might suggest that engineers and designers can leverage human heuristics and prior experiences to gain insights that are not easily captured by current computational tools. 

Interestingly, participants' accuracy decreased when both the visual rendering and the numerical performance data were provided, compared to the numerical performance data only condition. It seems that the added visual information changed the participants' minds, and they then made selections based on other factors. As in the visual rendering condition, participants may be getting additional information from the rendering that isn’t captured by the performance data. It could also mean the visual renderings are biasing people’s perceptions or causing design fixations.

\subsection{Designers prefer conventional designs with good performance}

In this study, participants showed clear preferences for conventional designs. Drone designs with conventional layouts and features have a significantly higher pick rate, namely, designs that are axisymmetric and in which all propellers and the main body lie on a single plane. Across both participant pools, participants showed a strong preference for conventional designs when provided with only visual renderings. Also, fewer participants picked the unusual designs when visual renderings became available, after transitioning from numerical performance data to the mixed modality. 

Moreover, there is a significant difference in pick rates for designs with conventional features (axisymmetric and one-plane) versus designs with atypical features (non-axisymmetric, non-planar, or off-plane) for the visual rendering only condition ($t(24) = 3.103, p = 0.014, \text{Cohen's } d = 1.458$). Participants picked more conventional axisymmetric and one-plane designs significantly more often than the other designs. Furthermore, the axisymmetricity seems to be the dominant factor, as there is a significant difference in pick rates for axisymmetric designs versus non-axisymmetric designs for the visual rendering only condition ($t(24) = 2.512, p = 0.023, \text{Cohen's } d = 0.866$). \textbf{Participants prefer axisymmetric designs.} 

\begin{figure}[H]
    \centering
    \includegraphics[width=0.5\linewidth]{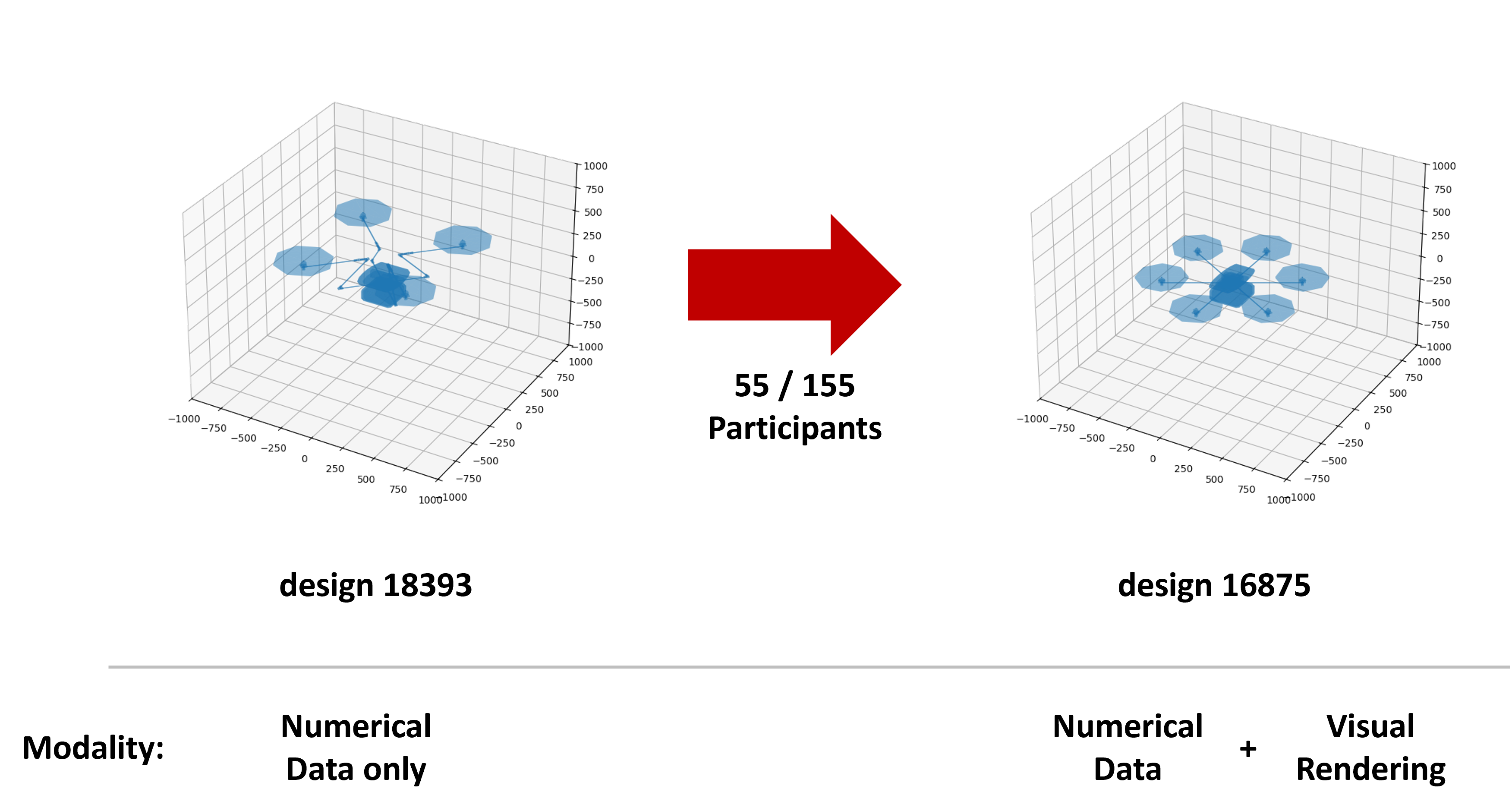}
    \caption{Changes in participants' choice of design for Design Problem with 2 options when the design modality changed from numerical performance data to visual rendering + numerical performance data. }
    \label{fig:transition_example_1}
\end{figure}

\begin{figure}[H]
    \centering
    \includegraphics[width=0.5\linewidth]{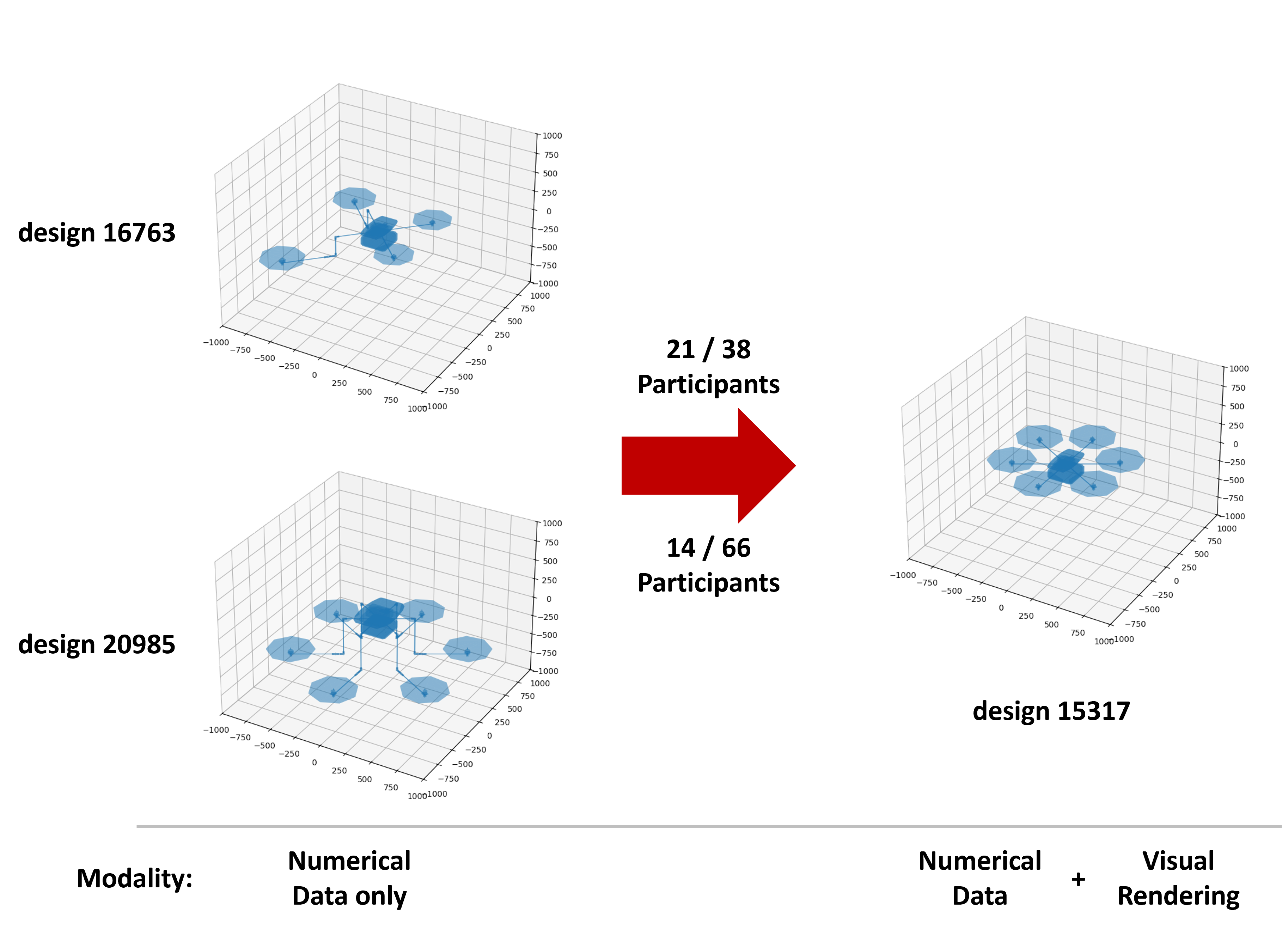}
    \caption{Changes in participants' choice of design for Design Problem with 16 options when the design modality changed from numerical performance data to visual rendering + numerical performance data. }
    \label{fig:transition_example_2}
\end{figure}

Changes in design choices across modalities further illustrate this preference. When the design modality changed from numerical performance data to the mixed modality, a noticeable number of participants moved away from the "unorthodox" designs with better performance toward designs with more conventional design features. This phenomenon is best illustrated with the first design problem with 2 options, where 55 out of 155 participants who chose the "unorthodox" design (design 18393) when shown only numerical performance data moved away and switched to the more conventional design 16875 when presented with both visual rendering and numerical performance data (Figure~\ref{fig:transition_example_1}). Similarly, in the third design problem with 16 options, 21 and 14 out of 38 and 66 participants, who chose the more unusual design 16763 and design 20985, two of the most picked designs, when provided only numerical performance data, switched to the more conventional design 15317 instead when presented with both visual rendering and numerical performance data (Figure~\ref{fig:transition_example_2}). These transitions clearly show participants' preference for conventional design features regardless of performance data. 

Participants' textual responses for their reasons for the design choices also reflect their preferences for axisymmetric and "reasonable-looking" designs. One participant in Study 2A said they \say{feel that symmetrical designs have a big impact on the hover time, so asymmetrical design D, design E, and design H were immediately ruled out} [Study 2A, P\#57]. A participant in Study 2A also ruled out asymmetrical designs as \say{options have large asymmetry would make maintenance harder and more expensive due to specialized parts}[Study 2A, P\#19]. Participant \#54 in Study 2B explained their design choice as \say{the rendering seems reasonable}. Another participant explained their preference for symmetrical design, as \say{symmetrical design would make manufacturing easier} [Study 2A, P\#26]. Reliability can be another factor that drives the preference for symmetry, as one participant said \say{Design D combines symmetry with rotors close to the center of mass to create a reliable design} [Study 2A, P\#18].

One participant in Study 2B clearly showed their preferences for "reasonable-looking" and good-performing design in their textual response, as they state \say{among the designs that look reasonable, Design K has the best statistics with a lower cost, and would probably work best for this scenario} [Study 2B, P\#27]. They also doubted the designs with unconventional features despite the better performance data, saying \say{while Design C has the best statistics, I don't think it would actually work in real life} [Study 2B, P\#27]. Interestingly, there are also participants who have higher faith in the numerical data, \say{even though A looks strange, it outperforms B in every metric. There is a chance that its unorthodox design lends it some unique advantages} [Study 2B, P\#08]. 

\textbf{Overall, participants prioritize designs with conventional features, and they prefer the better-performing designs among those with conventional features.} Such design heuristics can be helpful for participants to identify the good performers, considering both simulation results and real-life scenarios. However, it can also mean that the participants are unknowingly or unintentionally omitting good-performing designs with atypical design features, leading to potential design fixation. Our work provides empirical evidence and insights into the prior work of Demirel et al. \cite{demirel_human-centered_2024}, showing how a human-centered generative design framework can be improved by choosing appropriate representations to capture and incorporate human factors and preferences.

\subsection{Designer’s judgment ability is weakened when presented with a large number of options}

Two-way ANOVA results on the interaction effect between the design modalities and the number of design options show that the set size of design options can affect participants' accuracy in identifying optimal designs. Since all three design problems have a single optimal design (based on Pareto-TOPSIS), we compared the participants' accuracy in selecting the optimal design across design problems with different numbers of design options. We found that, across all three modality conditions, participants were less accurate in selecting the optimal design in problems with 8 or 16 design options than in problems with only 2 options. This may suggest that with more potential design solutions provided, participants’ ability to determine the optimal design is weakened. 

One possibility is that with fewer options shown, participants' mental capacity is sufficient to examine solutions and perform pairwise comparisons throughout and simultaneously, leading to more comprehensive design interpretations. When a large number of candidate solutions are provided, e.g., 8 or 16 options, the amount of design information to consider exceeds participants' cognitive capacity. Participants might be overwhelmed by the potential options and pairwise comparisons when faced with complex and challenging design requirements. Their abilities and willingness to explore more designs or more novel designs can be hindered. This finding resonates with research in working memory, which suggests that the capacity of human working memory is limited at a given time and is typically limited to 4 objects for \textbf{visual working memory and short-term memory} \cite{luck_capacity_1997, cowan_magical_2001}, although the classic research in working memory by George Armitage Miller argued a larger 7 objects with a range of plus or minus 2 \cite{miller_magical_1956}. Future research should investigate this issue further, along with a more explicit examination of the impact of the number of design options offered on the designer's decision-making.

\subsection{Implications and recommendations for generative design systems}

The findings from this work inform the future development and implementation of generative engineering design tools and design comparison tools used with generative AI systems. Specifically, we have three recommendations. 

\begin{enumerate}
    \item Consider limiting the number of options presented at the same time for user selection or comparison. In this work, we found that designers’ judgment ability is weakened when presented with a large number of options. Prior research on working memory suggests the limit is around 7 options \cite{miller_magical_1956}, significantly fewer than the hundreds or thousands of options that generative design systems can produce. More research is needed to determine task-specific bounds for engineering design where designers are seeking both inspiration and optimality. 

    \item Consider allowing the user to control whether to show or hide visual representations (e.g., geometric representations of physical products or sketches) when presenting design solutions. In this work, we found that design modalities affect design decision-making. Allowing modality toggling gives the user the option to either conform to the paradigm or break the paradigm when exploring the design space \cite{silk_incremental_2019}, depending on their needs and objectives (e.g., novelty vs well-accepted conventions). 

    \item Consider providing more tools or promoting a process that helps designers first avoid their biases (e.g., working only with numerical data), then bring in more information (e.g., adding visual design renderings). In this work, we observed that the human-AI collaboration loop might be changed by how and when design information is presented. Many participants filtered out novel or unconventional designs once they saw visual renderings, and then picked the optimal designs within the reduced solution set. Facilitating tools or processes for bias mitigation could improve the human-AI collaboration loop. 
    
\end{enumerate}

\subsection{Limitations and future works}

This work faces several limitations. First, we evaluated design optimality using a physics-based simulator. Such a simulator is accurate regarding the physics phenomena and drone design features that are modeled, but it can neglect meticulous factors and lead to inaccurate evaluations in real-life scenarios. Therefore, the simulation results and performance evaluation in this work may not be perfectly accurate for the tasks described in the design problems, potentially affecting our evaluation of participants' ability to select optimal designs. 

Second, this study examined a domain-specific design problem with relatively high requirements for domain knowledge. Participants with prior UAV design, manufacturing, or flying experience will likely have a more comprehensive heuristic and can demonstrate a stronger ability to evaluate UAV designs than those without it, potentially affecting the results. Future work should further investigate the impact of design modalities on more general design problems and other technical domains. 

Third, all participants in this study are current college students. Though it is a representative group to demonstrate the impact of different design modalities and experiment conditions, the participants may lack the field experience for some of the realistic technical challenges reflected in the design problems, which sometimes can only be obtained through years of industry-specific work experience. Future work should consider professional engineers and designers, and further investigate the impact of expertise. 

Fourth, we limited the additional information included in the interactive 3D renderings of the UAV designs used in the study (i.e., no context, functional annotations, or component animation) to accentuate and highlight the effect of design representation modality. However, this might deflate their usefulness and realism, as real-world design renderings typically include reference information that could mitigate bias. Meanwhile, we presented the numerical design performance data with tables along with spider plots, which are regarded as common practice in the aerospace industry. Since extensive research has shown that different styles of data visualization can affect the interpretation of visual information, future work should also examine the impact of different design visualization methods on design decision-making \cite{perer_integrating_2008, correll_value-suppressing_2018, moritz_formalizing_2019}. 

Moreover, we observed that the number of design options affected participants' ability to identify optimal designs, but option set size was not manipulated as a primary independent variable. It is worth investigating the impact of design option set size on people's design decision-making more directly in future studies. Besides, this work focuses more on the outcomes of design decision-making rather than the process. Future work should further and more comprehensively study the influence of different design modalities on designers' decision-making behavior and rationale. 

In this work, our goal was to isolate individual judgment under controlled conditions, rather than recreate the unwieldy full complexity of engineering design collaboration. Therefore, we focused on individual, simplified design tasks simulated via an online survey tool (Qualtrics), rather than collaborative, sophisticated design tasks. Also, although no hard limit was imposed, most participants finished the study within 40 minutes. Such a time span may not fully and accurately represent the extensive nature of real-world design challenges. However, the findings of this work still have real-world relevance, as early-stage filtering and selection of design concepts is often done individually. Also, individual design biases, if systematically rooted in design modality and the human-AI collaboration loop, can cascade into downstream team decisions. Therefore, the isolated individual judgment studied in this work is meaningful outside controlled conditions. Our findings serve as an explanatory step in the research on human-AI collaboration, and as a starting point for new research on cognitive mechanisms in the use of generative design tools. Future work should further investigate design decision-making and human-AI collaboration in more complex, extensive, and collaborative work settings.

\section{Conclusion}

This work examines the impact of different design modalities on the design decision-making process of engineers and designers. More specifically, we investigate whether design visualization affects engineers' and designers' ability to select the optimal designs from a list of AI-generated design ideas. We found that different design modalities do affect engineers' decision-making when using AI-powered generative design tools. For the participants in this study, providing only the numerical design performance data leads to the best accuracy in selecting the most optimal design, while only seeing design renderings provides marginal help in selecting optimal designs. We found that presenting both the numerical design performance data and the design rendering results in worse accuracy compared to seeing the numerical performance data alone, suggesting that engineers can leverage design heuristics when providing the design visualization, or, alternatively, the presence of design visualization can induce design biases and fixations. In addition, we found that the participants generally prefer the best-performing designs \textit{as long as} those designs possess traditional and symmetrical appearances. Also, we note that the number of design options provided affects people's ability to choose optimal designs in this study, and a large number of design options can overwhelm engineers and lead to suboptimal design choices. This work deepens our understanding of how people interact with AI design systems that generate many options with the goal of giving novel and optimal designs. Our findings on the impacts of design modality on decision making can guide the future development and implementation of generative engineering design tools and design comparison tools used with generative AI systems.


\section*{Acknowledgments}
Thanks are due to Dominik Moritz and Adam Perer for their indispensable assistance and feedback on the experimental design of this work. The authors are grateful to researchers at SRI International Research Lab and Southwest Research Institute for their generous resources and help in developing the UAV flight simulator. The authors are also grateful to Allison Fisher and Ranald Engelbeck for their invaluable feedback and tremendous help in recruiting participants to make this work possible.

\bibliographystyle{plainnat}
\bibliography{references}

\appendix

\newpage
\section{Design Problem}

Imagine you are a lead design engineer working in an engineering consulting company designing UAVs tailored to customer needs. Your team uses an AI-powered automatic design generation system to help ideate and create initial design solutions. These solutions have been tested in a newly developed computer simulation environment. Your simulation teams assure you that the simulator is the best in the business and that the simulated performance data is accurate. However, as an engineer, you still need to use your expertise and engineering knowledge to inspect and evaluate the designs while considering real-world scenarios and pick the best design for further physical testing and validation before delivering it to the customers.

Here is your task: 

Pennsylvania’s fire department is using UAVs to monitor wildfires. The UAV should have great hover time (maintain target position with minimal deviation) and carry at least 10 kg (22.0 lbs) of monitoring equipment, including RGB cameras, IR cameras, and other sensors. The UAV must rise to a designated height at a fast vertical lift speed. Without sacrificing hover time, more carrying capacity is desirable for carrying more equipment for better coverage. The UAV must also fly steadily and sustain cross-wind and other potential environmental hazards for safety reasons. There is no requirement for maximum travel distance and travel speed.

Important Design Information:
The best design refers to the design you deem most optimal considering all factors.
Hover time is not air time. Hover time means the UAV is hovering at the target position in mid-air.
The designs are not presented in any particular order. There is no correlation between the designs' order and their performance. The designs are only presented in the order in which they are generated by the AI. 
This is an AI-generative system. The design may or may not work in real life.
Please use your engineering knowledge and judgment. Consider all factors, including external ones, that may not have been considered and simulated by the AI.

\newpage
\section{UAV Designs Used in the Study with Design Features}

\begin{longtable}{|l|l|l|}
\hline
 List                & Design       & Rendering \\
\hline
\endhead
 List 1 - 2 options  & \textbf{design\_18393*} & \includegraphics[width=3cm]{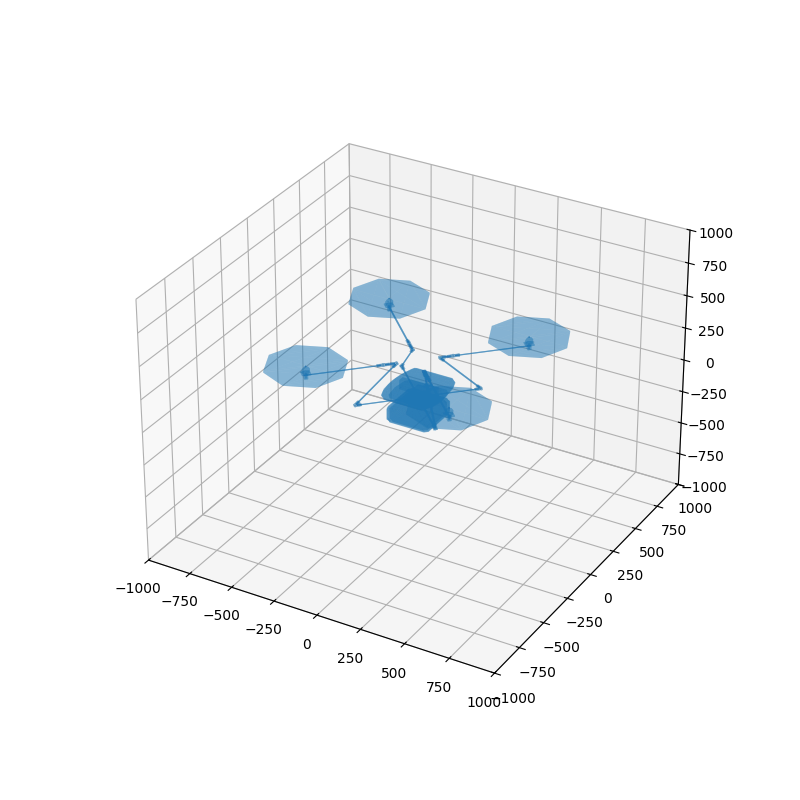} \\
 \hline
 List 1 - 2 options  & design\_16875 & \includegraphics[width=3cm]{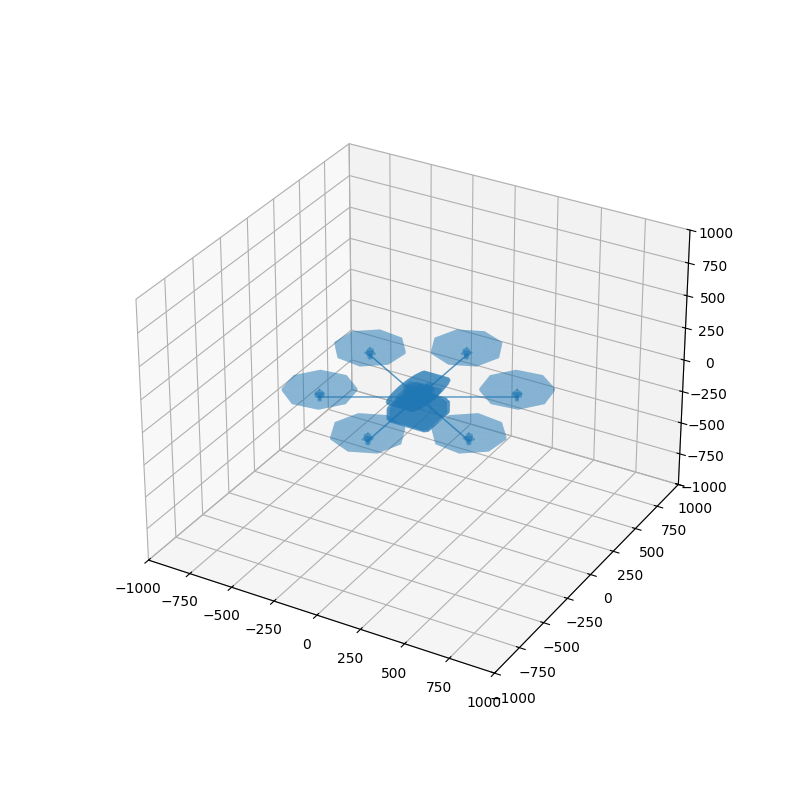} \\
\hline

\caption{Design Rendering Table - List 1 (2 options). * indicates the Pareto-TOPSIS optimal design. }
\label{tab:drone_rendering_1}
\end{longtable}

\begin{landscape}
\begin{longtable}{|l|l|l|l|l|l|}
\hline
 List                & Design       & Rendering  &  List                & Design       & Rendering\\
\hline
\endhead
List 2 - 8 options  & design\_2986  & \includegraphics[width=3cm]{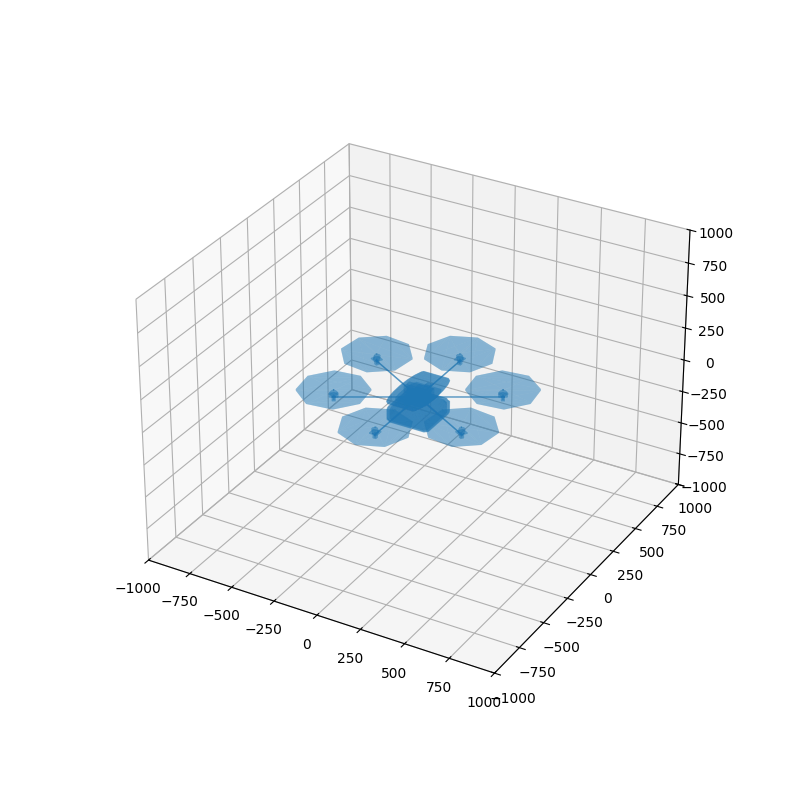} & List 2 - 8 options  & design\_25944 & \includegraphics[width=3cm]{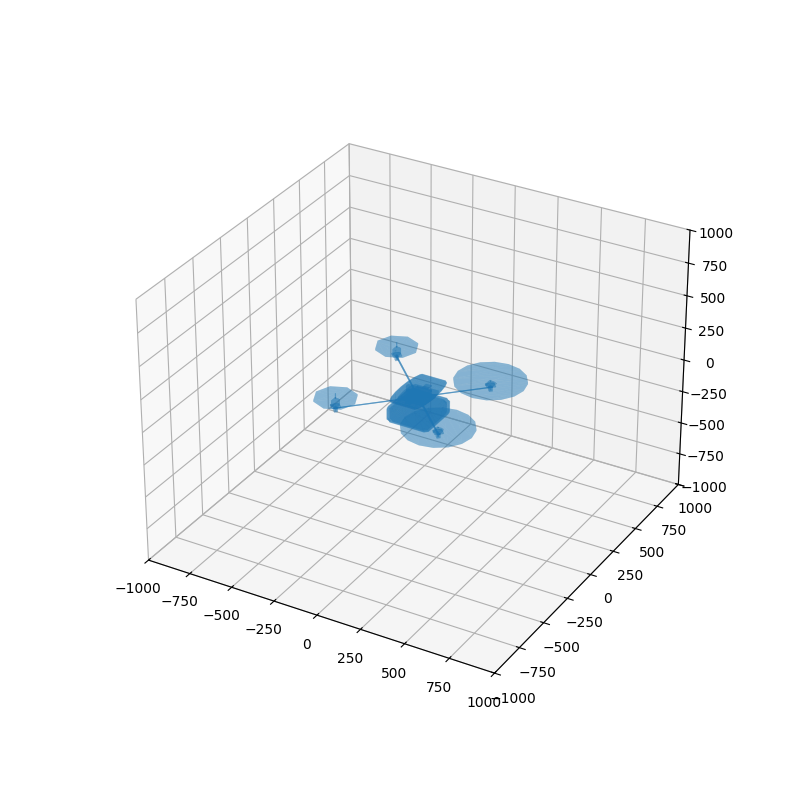} \\
\hline
List 2 - 8 options  & design\_9510  & \includegraphics[width=3cm]{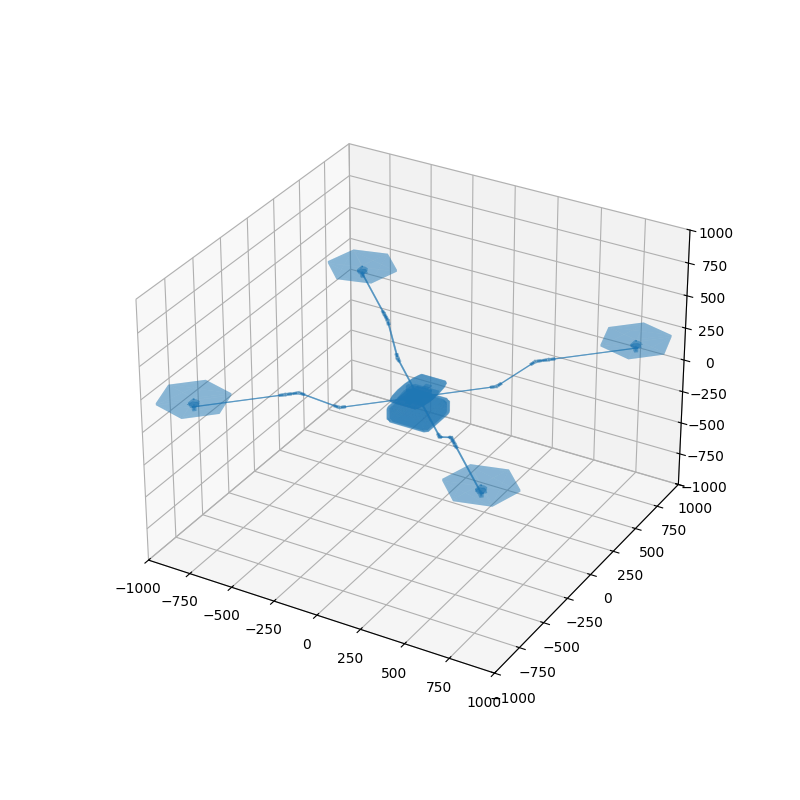} & List 2 - 8 options  & design\_27150 & \includegraphics[width=3cm]{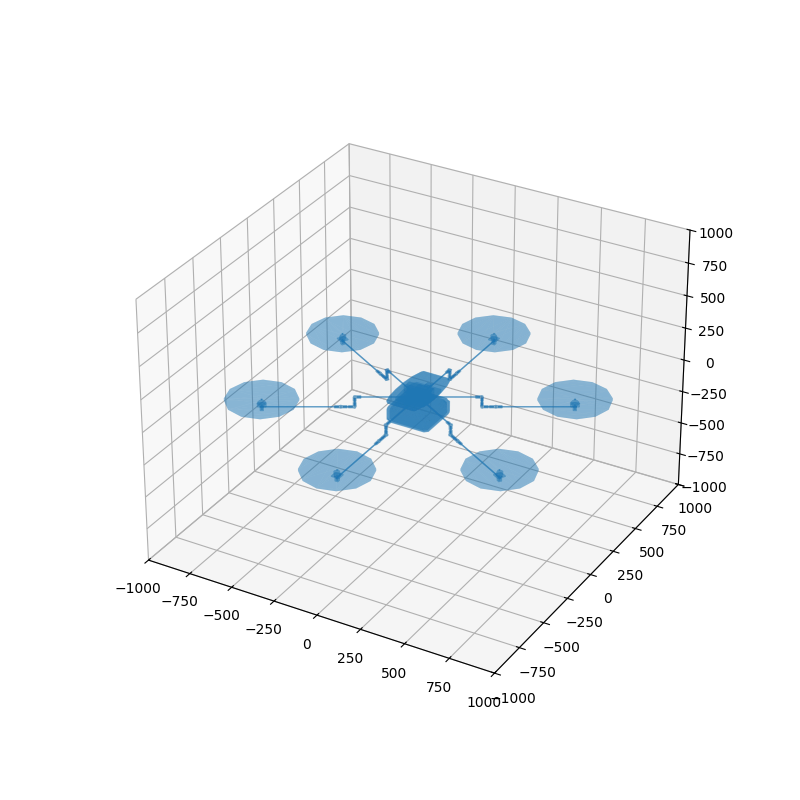} \\
\hline
List 2 - 8 options  & \textbf{design\_20155*} & \includegraphics[width=3cm]{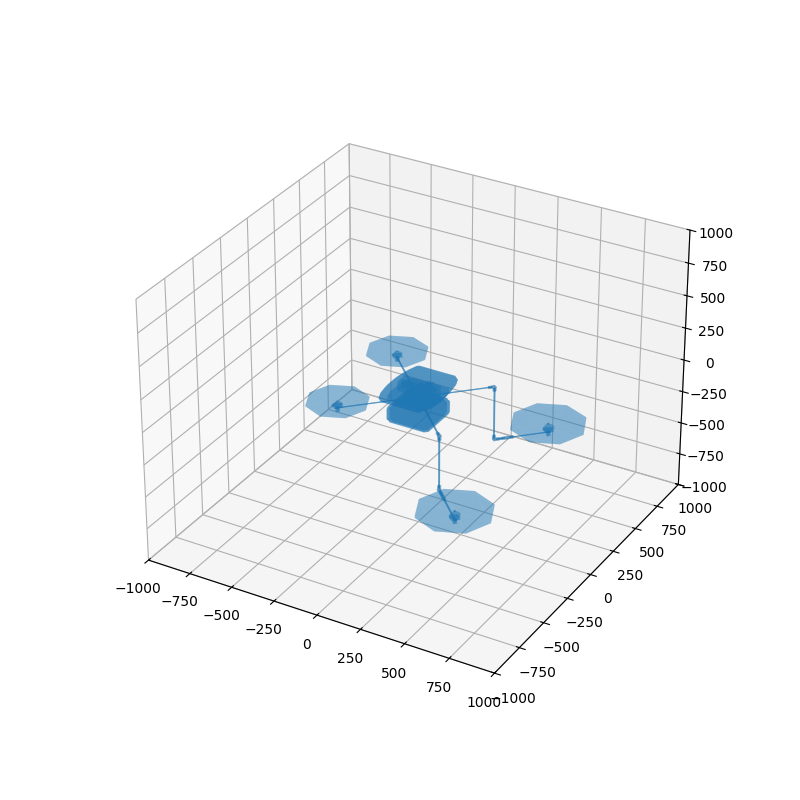} & List 2 - 8 options  & design\_27450 & \includegraphics[width=3cm]{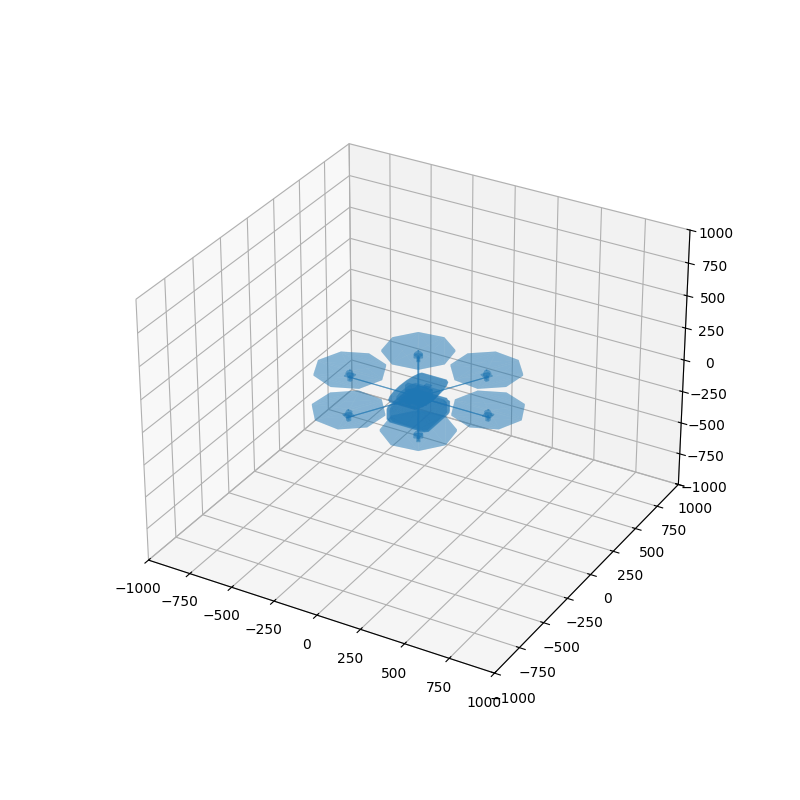} \\
\hline
List 2 - 8 options  & design\_25139 & \includegraphics[width=3cm]{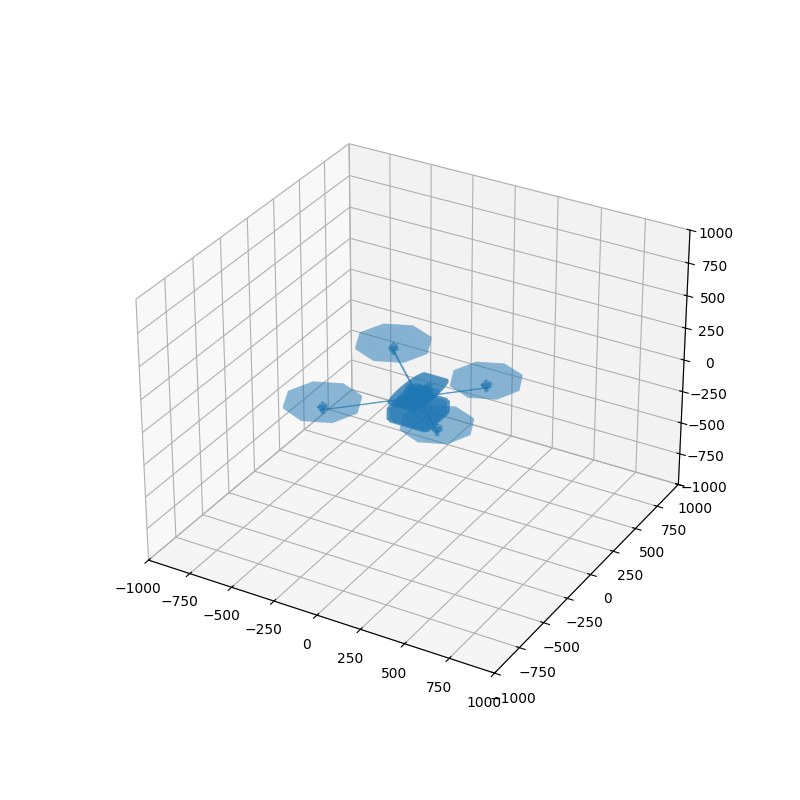} & List 2 - 8 options  & design\_27604 & \includegraphics[width=3cm]{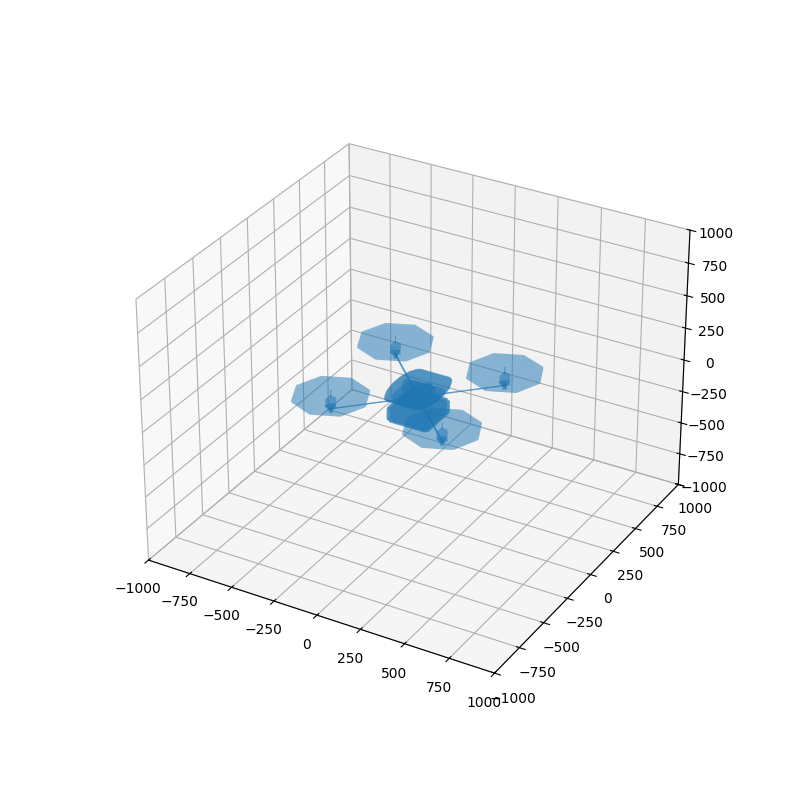} \\
\hline
\caption{Design Rendering Table - List 2 (8 options). * indicates the Pareto-TOPSIS optimal design. }
\label{tab:drone_rendering_2}
\end{longtable}

\end{landscape}

\begin{landscape}
\begin{longtable}{|l|l|l|l|l|l|}
\hline
 List                & Design       & Rendering  &  List                & Design       & Rendering\\
\hline
\endhead

List 3 - 16 options & design\_14962 & \includegraphics[width=3cm]{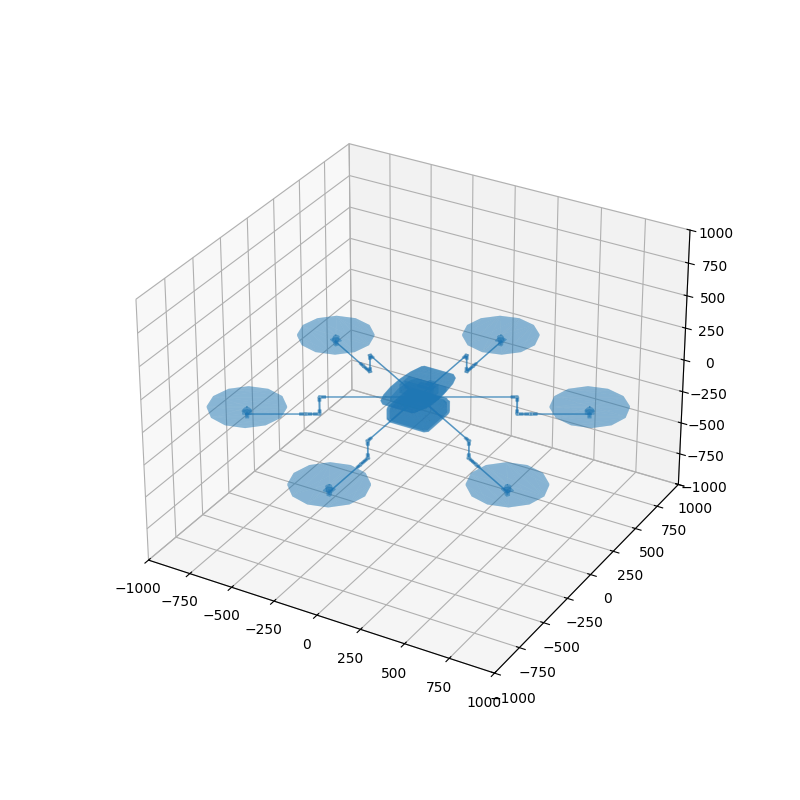} & List 3 - 16 options & design\_18802 & \includegraphics[width=3cm]{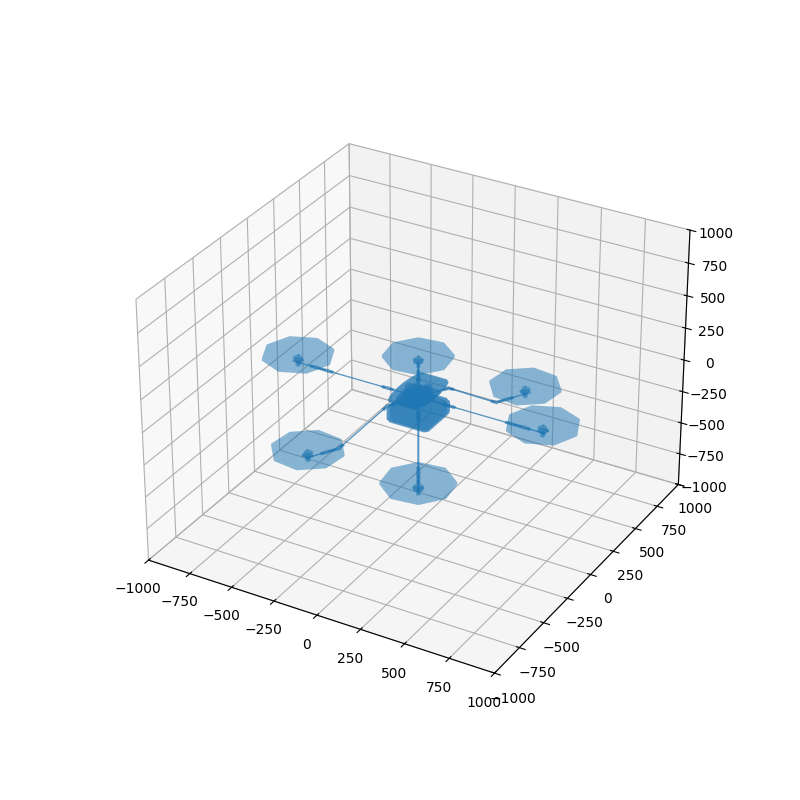} \\
\hline
List 3 - 16 options & design\_15317 & \includegraphics[width=3cm]{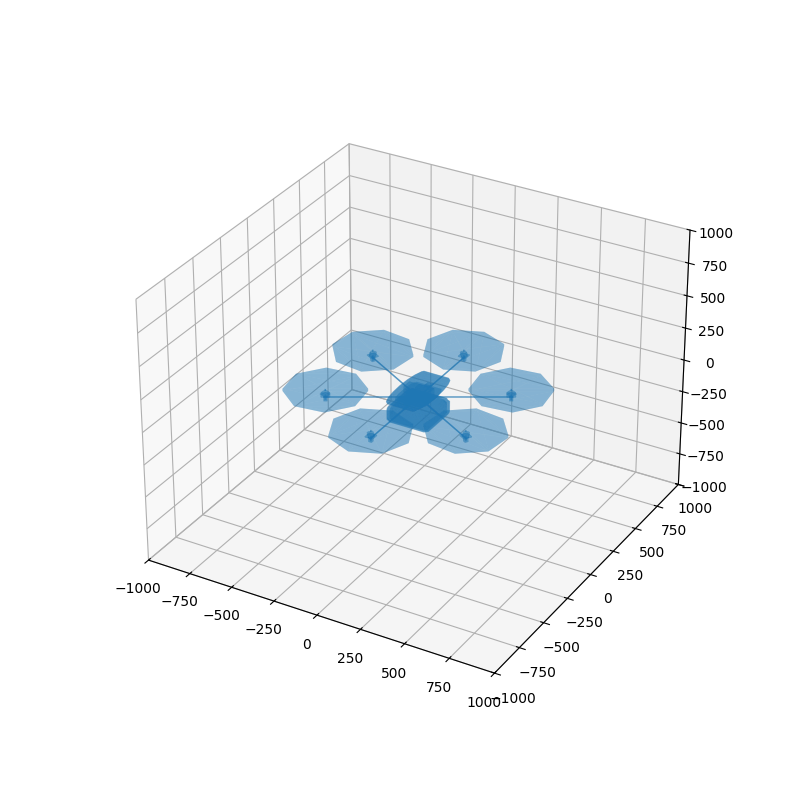} & List 3 - 16 options & design\_18914 & \includegraphics[width=3cm]{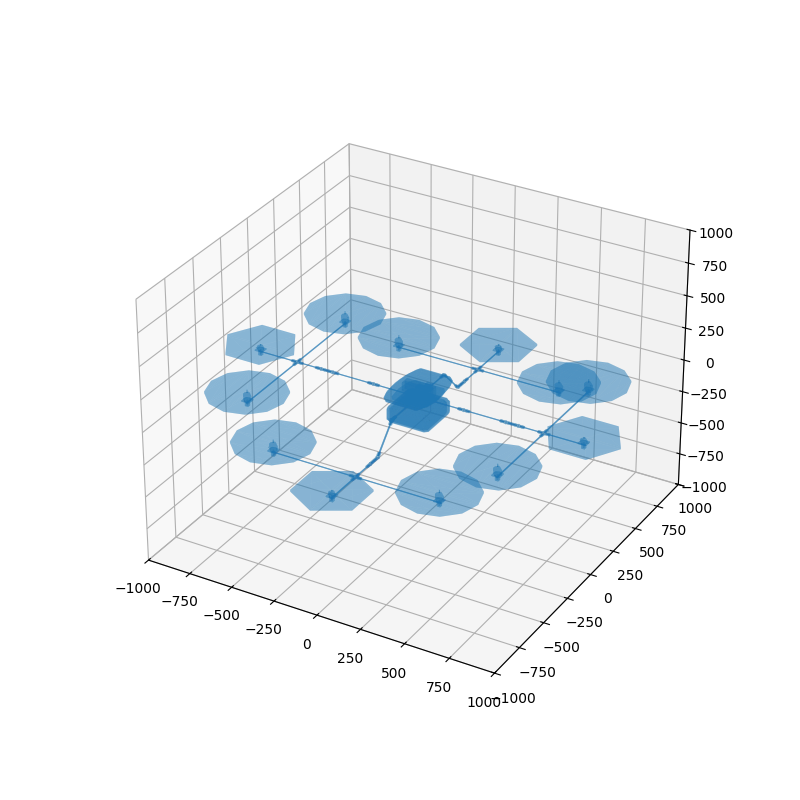} \\
\hline
List 3 - 16 options & design\_16763 & \includegraphics[width=3cm]{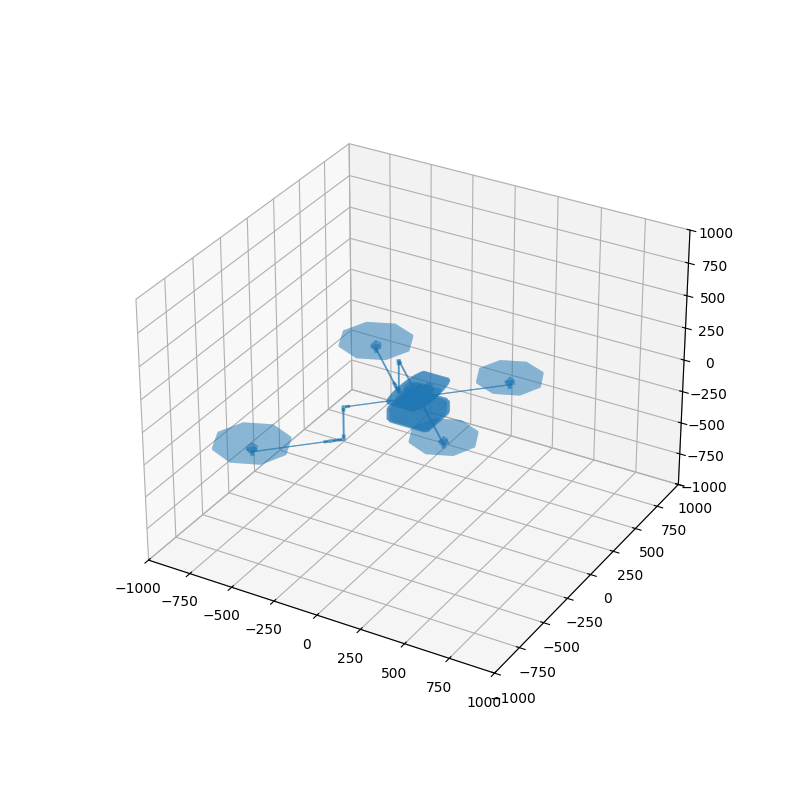} & List 3 - 16 options & design\_18952 & \includegraphics[width=3cm]{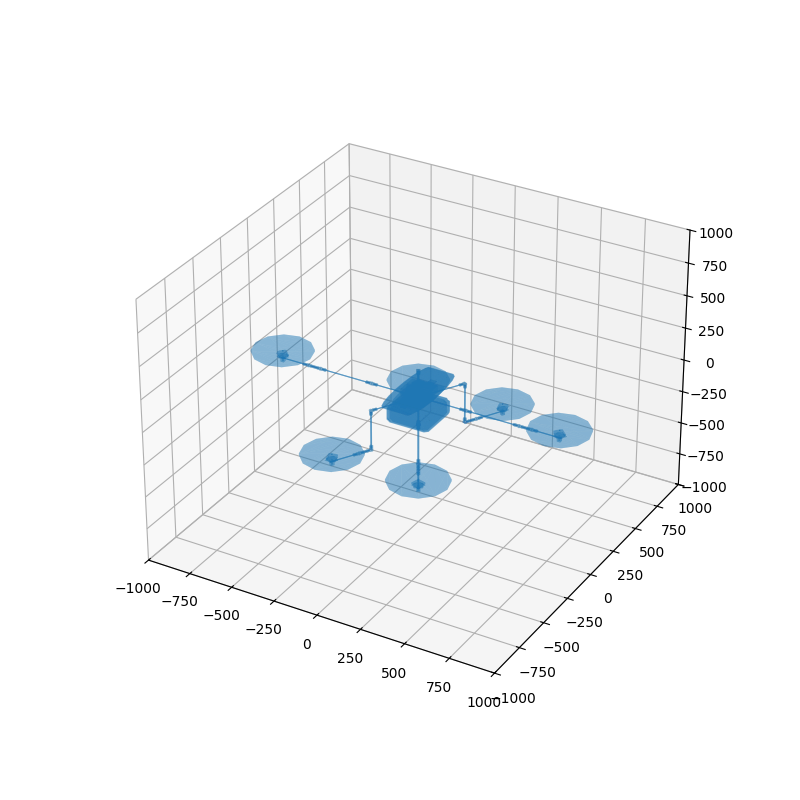} \\
\hline
List 3 - 16 options & design\_18368 & \includegraphics[width=3cm]{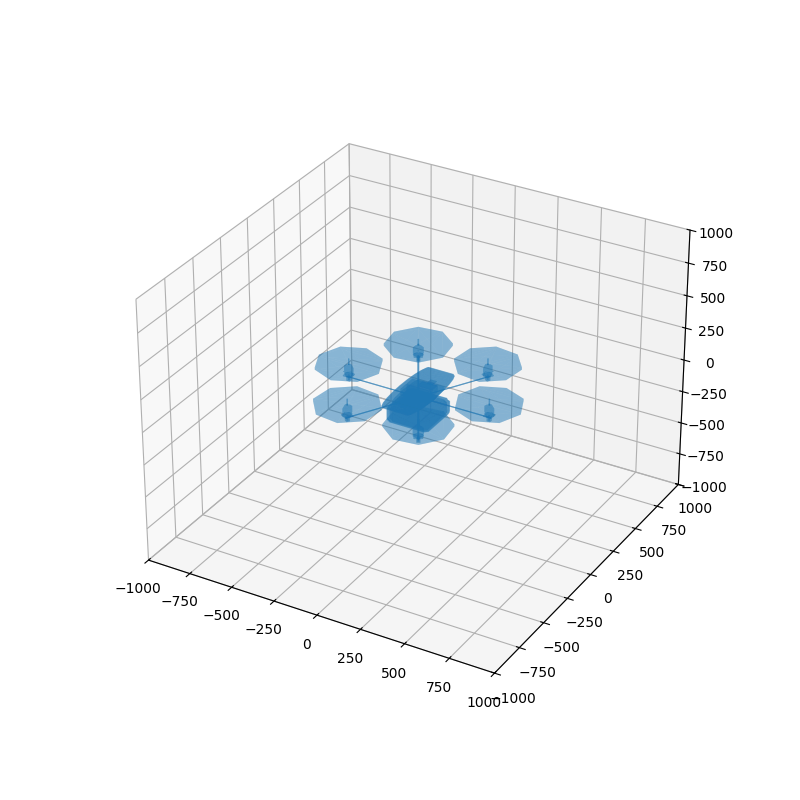} & List 3 - 16 options & design\_19051 & \includegraphics[width=3cm]{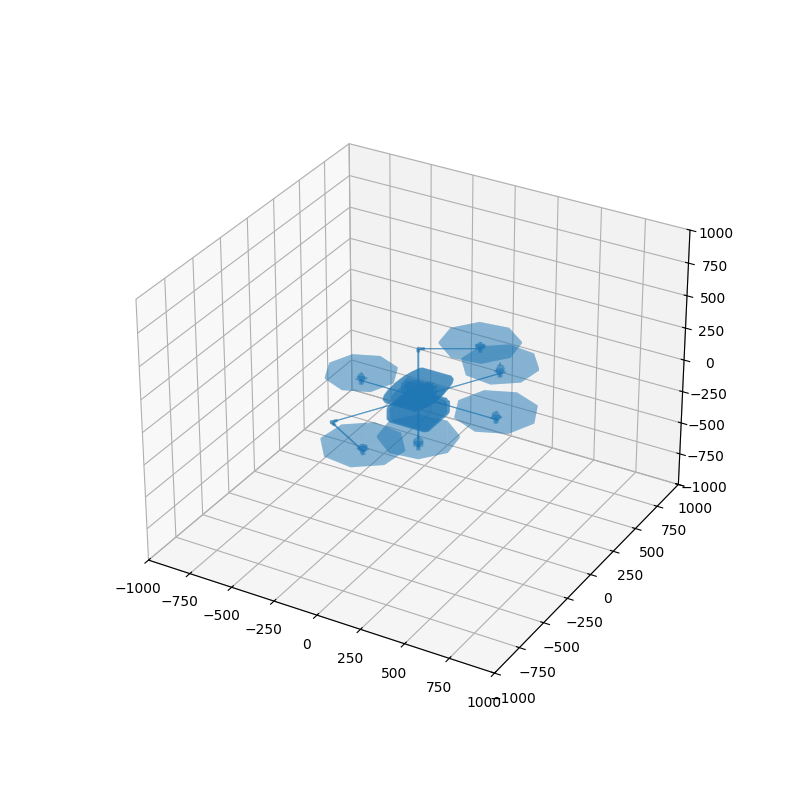} \\
\hline

\caption{Design Rendering Table - List 3 (16 options). * indicates the Pareto-TOPSIS optimal design. }
\label{tab:drone_rendering_3}
\end{longtable}

\end{landscape}

\begin{landscape}
\begin{longtable}{|l|l|l|l|l|l|}
\hline
 List                & Design       & Rendering  &  List                & Design       & Rendering\\
\hline
\endhead

List 3 - 16 options & design\_20320 & \includegraphics[width=3cm]{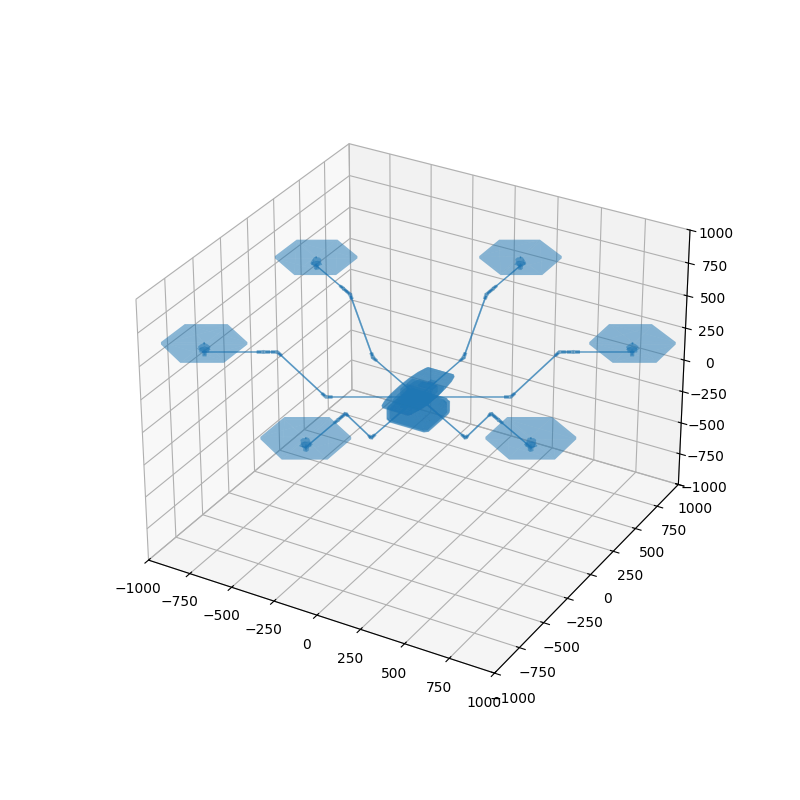} & List 3 - 16 options & design\_26003 & \includegraphics[width=3cm]{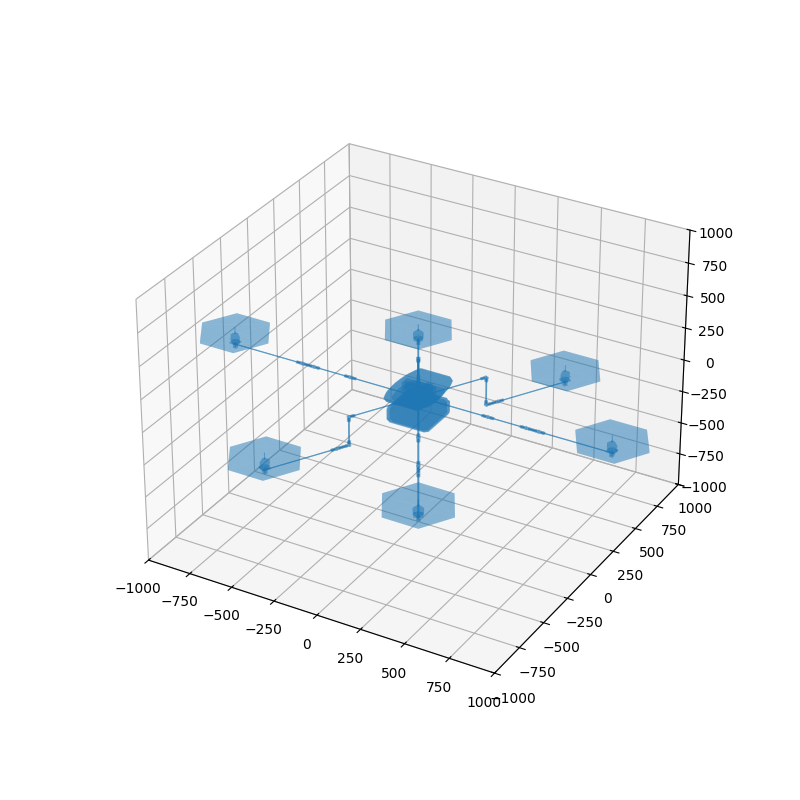} \\
\hline
List 3 - 16 options & \textbf{design\_20985*} & \includegraphics[width=3cm]{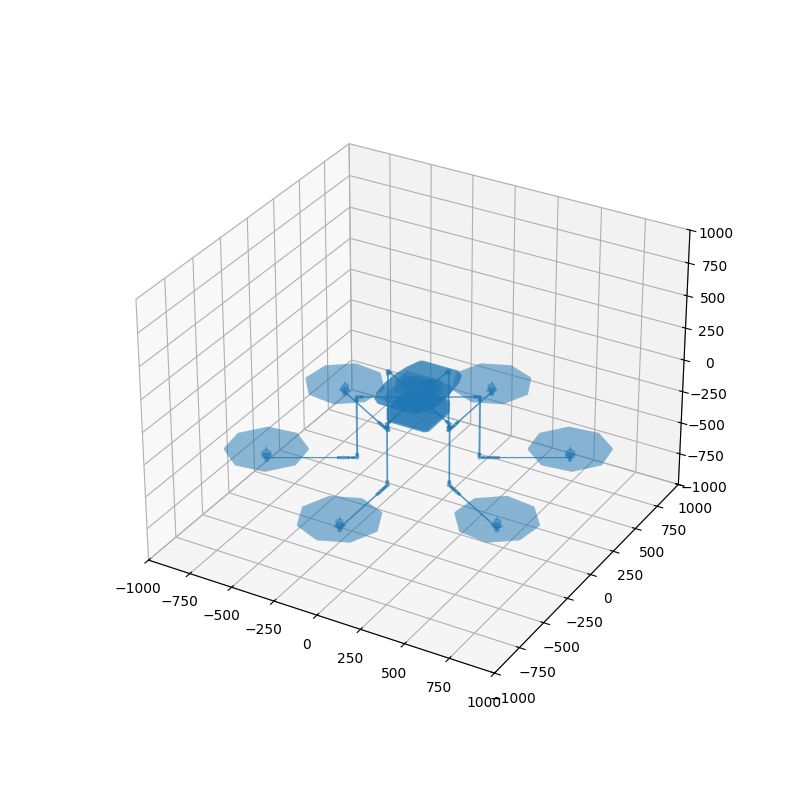} & List 3 - 16 options & design\_26842 & \includegraphics[width=3cm]{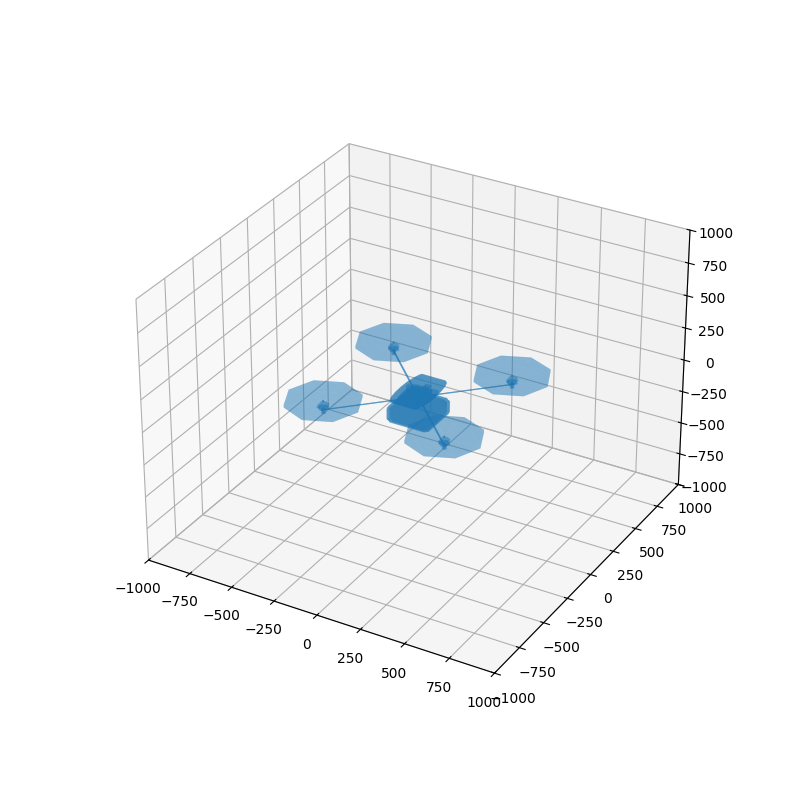} \\
\hline
List 3 - 16 options & design\_24246 & \includegraphics[width=3cm]{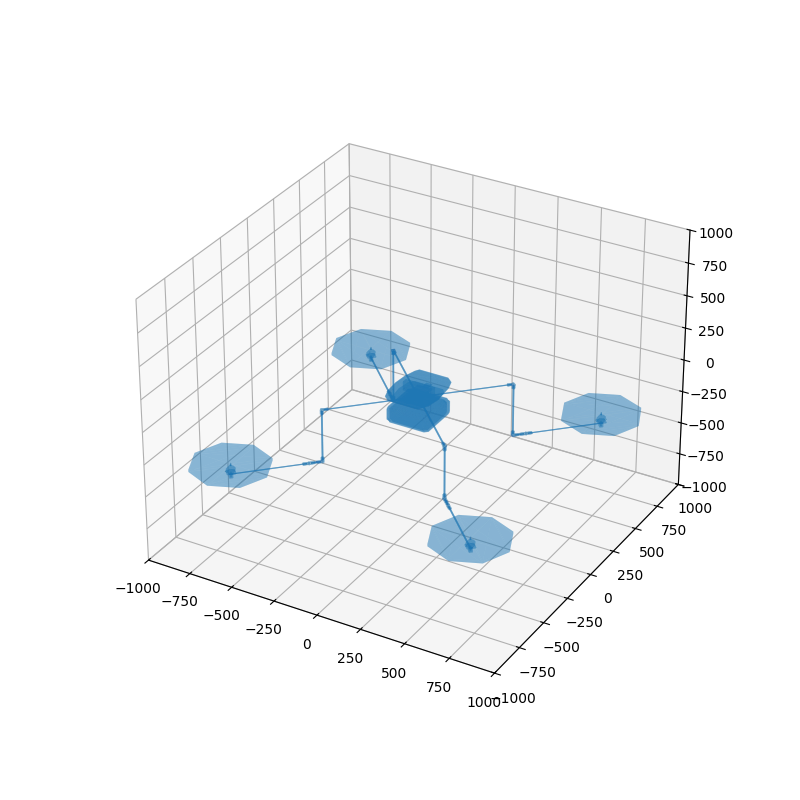} & List 3 - 16 options & design\_27369 & \includegraphics[width=3cm]{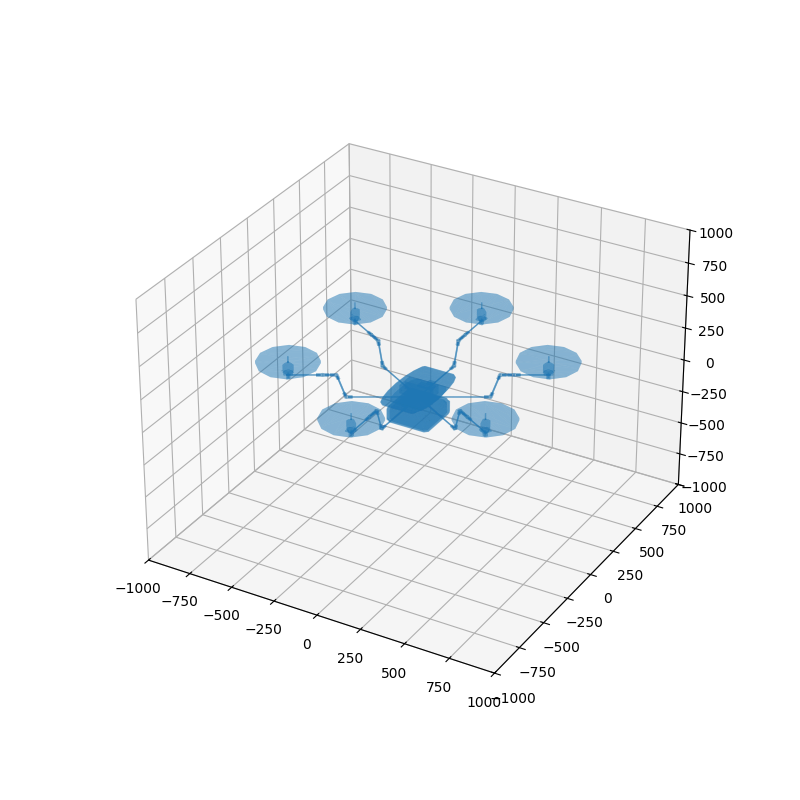} \\
\hline
List 3 - 16 options & design\_25633 & \includegraphics[width=3cm]{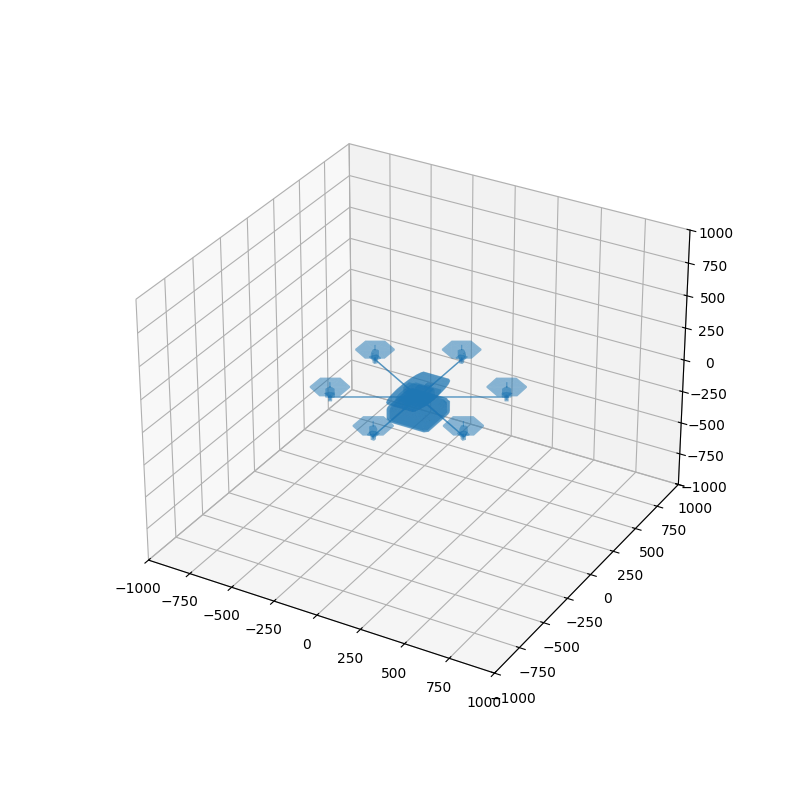} & List 3 - 16 options & design\_27901 & \includegraphics[width=3cm]{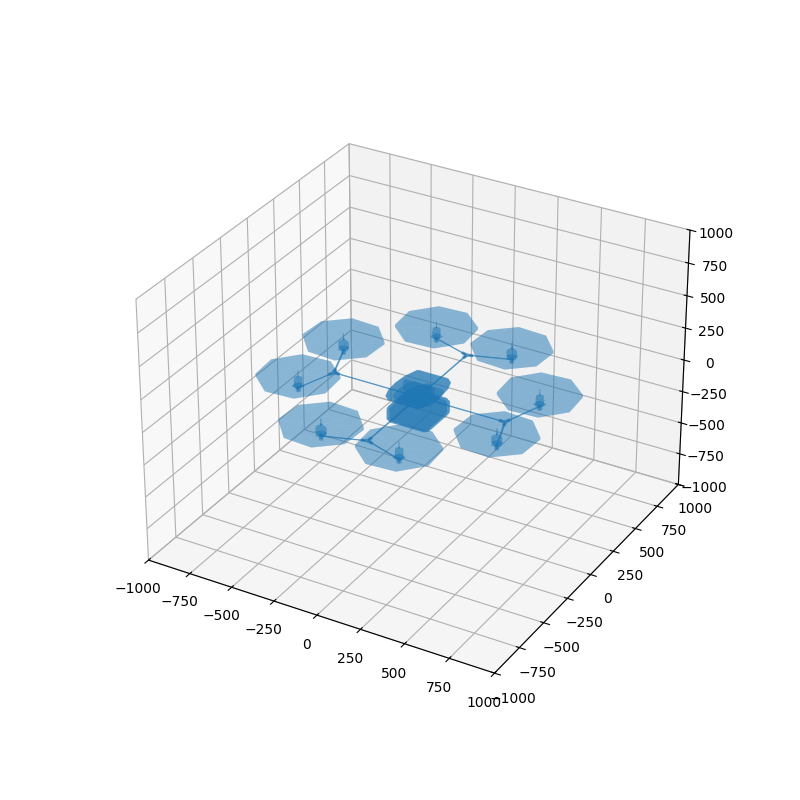} \\
\hline
\caption*{Table 7 (Continued.) Design Rendering Table - List 3 (16 options) (Continued). * indicates the Pareto-TOPSIS optimal design. }
\end{longtable}

\end{landscape}

\begin{landscape}
\begin{longtable}{llccccccc}
\hline
\makecell{List} & \makecell{Design} & \makecell{Hover Time\\(seconds)} & \makecell{Max Travel\\Distance (meters)} & \makecell{Mass\\($kg$)} & \makecell{Max Speed\\($m/s$)} & \makecell{Battery\\Voltage ($V$)} & \makecell{Total Cost\\(\$)} &   \makecell{Max Lift\\($kg$)} \\
\hline
\endhead
 List 1 - 2 options  & \textbf{design\_18393*} &       428.67 &        7189.98 &       4.98 &          25 &              22.2 &       657.42 &      50.24 \\
 List 1 - 2 options  & design\_16875 &       211.81 &        3477.01 &       4.82 &          22 &              11.1 &       844.66 &      27.09 \\
\hline
 List 2 - 8 options  & design\_2986  &       246.01 &        2596.37 &       3.99 &          11 &              11.1 &       667.50 &      13.88  \\
 List 2 - 8 options  & design\_9510  &       177.91 &        2627.52 &       3.63 &          18 &              11.1 &       482.14 &      10.72 \\
 List 2 - 8 options  & \textbf{design\_20155*} &       306.71 &        5866.26 &       6.24 &          21 &              22.2 &       995.12 &      45.00 \\
 List 2 - 8 options  & design\_25139 &       181.70 &        2329.96 &       3.71 &          14 &              11.1 &       572.78 &      11.24 \\
 List 2 - 8 options  & design\_25944 &       151.24 &        2004.73 &       3.15 &          14 &              11.1 &       638.42 &       9.61 \\
 List 2 - 8 options  & design\_27150 &       203.33 &        2782.17 &       4.74 &          16 &              11.1 &       691.50 &      14.70 \\
 List 2 - 8 options  & design\_27450 &       308.99 &        3001.63 &       3.75 &          10 &              11.1 &       622.88 &      10.79 \\
 List 2 - 8 options  & design\_27604 &       230.78 &        2468.81 &       6.28 &          11 &              22.2 &      1453.79 &      18.42 \\
\hline
 List 3 - 16 options & design\_14962 &       207.87 &        3472.91 &       6.38 &          24 &              22.2 &       817.38 &      87.61 \\
 List 3 - 16 options & design\_15317 &       194.30 &        3096.53 &       4.70 &          30 &              22.2 &       806.02 &     154.35 \\
 List 3 - 16 options & design\_16763 &       209.68 &        3762.22 &       4.29 &          30 &              22.2 &       646.02 &      65.97 \\
 List 3 - 16 options & design\_18368 &        59.72 &         858.54 &       8.65 &          14 &              22.2 &      2218.34 &      81.42 \\
 List 3 - 16 options & design\_18802 &       145.98 &        2766.04 &       4.79 &          24 &              14.8 &       730.23 &      55.55 \\
 List 3 - 16 options & design\_18914 &       106.54 &        1592.01 &       8.85 &          15 &              11.1 &      1457.84 &      28.50 \\
 List 3 - 16 options & design\_18952 &       150.31 &        3144.53 &       6.93 &          22 &              22.2 &       880.04 &      41.56 \\
 List 3 - 16 options & design\_19051 &       191.56 &        3441.67 &       5.77 &          30 &              14.8 &       840.41 &      36.11 \\
 List 3 - 16 options & design\_20320 &       101.82 &        1998.72 &       7.29 &          36 &              22.2 &      1444.34 &     182.29 \\
 List 3 - 16 options & \textbf{design\_20985*} &       237.54 &        6337.14 &      10.78 &          30 &              22.2 &      1583.72 &     241.37 \\
 List 3 - 16 options & design\_24246 &       189.96 &        3116.33 &       4.86 &          35 &              14.8 &       539.76 &      32.02 \\
 List 3 - 16 options & design\_25633 &       114.83 &        2133.10 &       5.15 &          40 &              22.2 &      1377.69 &     165.03 \\
 List 3 - 16 options & design\_26003 &       105.14 &        2282.49 &       7.89 &          24 &              11.1 &      1932.02 &      29.78 \\
 List 3 - 16 options & design\_26842 &       102.81 &        1760.31 &       3.87 &          29 &              11.1 &       850.34 &      52.65 \\
 List 3 - 16 options & design\_27369 &       195.16 &        4241.97 &       9.64 &          32 &              22.2 &      2299.70 &      95.79 \\
 List 3 - 16 options & design\_27901 &       109.52 &        2016.81 &       8.32 &          23 &              11.1 &      2450.76 &      32.90 \\
\hline
\caption{Design Features Table - Design Information Explicitly Available to the Participants. * indicates the Pareto-TOPSIS optimal design. }
\label{tab:drone_feature_1}
\end{longtable}

\end{landscape}

\begin{landscape}
\begin{longtable}{lcccccccc}
\hline
\makecell{Design}       &   \makecell{Max Thrust\\($N$)} & \makecell{Effective Lift\\($kg$)} & \makecell{Propeller\\Area ($m^2$)} &   \makecell{Total Rod\\Length ($mm$)} &   \makecell{Rod Length to\\Area Ratio ($m^{-1}$)} &   \makecell{Number of\\Rods} &   \makecell{Number of\\Connectors} & \makecell{Number of\\Propellers} \\
\hline
\endhead
\textbf{design\_18393*} &  287.22 &   45.26 & 0.59 &           4172.23 &             7.13 &           12 &                  8 &  4 \\
design\_16875 &   85.85 &   22.27 & 0.68 &           2988.41 &             4.37 &            6 &                  0 &  6 \\
\hline
design\_2986  &   55.10 &    9.89 & 0.68 &           2562.89 &             3.75 &            6 &                  0 &  6 \\
design\_9510  &   58.12 &    7.09 & 0.46 &           4363.65 &             9.58 &           12 &                  8 &  4 \\
\textbf{design\_20155*} &  218.41 &   38.76 & 0.43 &           2829.59 &             6.58 &            8 &                  4 &  4 \\
design\_25139 &   57.50 &    7.53 & 0.49 &           1705.51 &             3.50 &            4 &                  0 &  4 \\
design\_25944 &   43.18 &    6.46 & 0.31 &           1617.17 &             5.22 &            4 &                  0 &  4 \\
design\_27150 &   79.16 &    9.96 & 0.68 &           4538.84 &             6.64 &           18 &                 12 &  6 \\
design\_27450 &   71.03 &    7.04 & 0.68 &           2413.34 &             3.53 &            6 &                  0 &  6 \\
design\_27604 &  200.79 &   12.14 & 0.52 &           1816.09 &             3.50 &            4 &                  0 &  4 \\
\hline
design\_14962 &  388.49 &   81.23 & 0.78 &           5331.75 &             6.86 &           18 &                 12 &  6 \\
design\_15317 &  538.61 &  149.65 & 0.88 &           2812.80 &             3.20 &            6 &                  0 &  6 \\
design\_16763 &  319.27 &  61.68 & 0.46 &           2968.60 &             6.49 &            8 &                  4 &  4 \\
design\_18368 &  171.79 &   72.77 & 0.60 &           2452.79 &             4.12 &            6 &                  0 &  6 \\
design\_18802 &  128.99 &   50.76 & 0.68 &           3608.00 &             5.28 &           18 &                 12 &  6 \\
design\_18914 &  223.89 &   19.65 & 1.69 &           7048.64 &             4.17 &           24 &                 12 & 12 \\
design\_18952 &  137.83 &   34.63 & 0.51 &           4165.84 &             8.11 &           18 &                 12 &  6 \\
design\_19051 &  231.15 &   30.34 & 0.81 &           3346.24 &             4.12 &            8 &                  2 &  6 \\
design\_20320 &  643.47 &  175.00 & 0.88 &           6852.84 &             7.81 &           18 &                 12 &  6 \\
\textbf{design\_20985*} & 1000.72 &  230.59 & 0.88 &           6833.34 &             7.78 &           18 &                 12 &  6 \\
design\_24246 &  250.13 &   27.16 & 0.59 &           5066.62 &             8.66 &           12 &                  8 &  4 \\
design\_25633 &  271.33 &  159.88 & 0.19 &           2671.68 &            13.75 &            6 &                  0 &  6 \\
design\_26003 &  171.97 &   21.89 & 0.78 &           5909.16 &             7.60 &           18 &                 12 &  6 \\
design\_26842 &  105.69 &   48.78 & 0.52 &           1971.66 &             3.80 &            4 &                  0 &  4 \\
design\_27369 &  138.71 &   86.15 & 0.51 &           4044.45 &             7.88 &           18 &                 12 &  6 \\
design\_27901 &  249.50 &   24.58 & 1.17 &           3798.75 &             3.25 &           12 &                  4 &  8 \\
\hline
\caption{Design Features Table - Design Information NOT Explicitly Available to the Participants. * indicates the Pareto-TOPSIS optimal design. }
\label{tab:drone_feature_2}
\end{longtable}

\end{landscape}

\begin{landscape}
\begin{longtable}{lrccccccc}
\hline
\makecell{Design}       &   \makecell{Identifying Feature}                        & \makecell{Non-\\axisymetric?} & \makecell{Non-\\planar?} & \makecell{Off-\\plane?} & \makecell{Pick\\\% V} & \makecell{Pick\\\% D} & \makecell{Pick\\\% M} & \makecell{Pick\\\% Overall} \\
\hline
\endhead
\textbf{design\_18393*} & off-plane quadcopter                       & NO       & NO       & YES          & 10.62\%      & 96.88\%      & 63.75\%      & 57.08\% \\
design\_16875 & hexacopter                                 & NO       & NO       & NO           & 89.38\%      & 3.12\%       & 36.25\%      & 42.92\% \\
\hline
design\_2986  & hexacopter                                 & NO       & NO       & NO           & 23.75\%      & 8.12\%       & 9.38\%       & 13.75\% \\
design\_9510  & off-plane quadcopter                       & NO       & NO       & YES          & 1.25\%       & 9.38\%       & 13.12\%      & 7.92\% \\
\textbf{design\_20155*} & non-axisym non-planar off-plane quadcopter & YES      & YES      & YES          & 1.25\%       & 47.50\%      & 30.00\%      & 26.25\% \\
design\_25139 & non-axisym quadcopter                      & YES      & NO       & NO           & 6.25\%       & 4.38\%       & 5.00\%       & 5.21\% \\
design\_25944 & non-axisym quadcopter                      & YES      & NO       & NO           & 3.75\%       & 0.00\%       & 1.25\%       & 1.67\% \\
design\_27150 & off-plane hexacopter                       & NO       & NO       & YES          & 5.62\%       & 5.00\%       & 6.25\%       & 5.62\% \\
design\_27450 & hexacopter                                 & NO       & NO       & NO           & 16.88\%      & 24.38\%      & 31.88\%      & 24.38\% \\
design\_27604 & quadcopter                                 & NO       & NO       & NO           & 41.25\%      & 1.25\%       & 3.12\%       & 15.21\% \\
\hline
design\_14962 & off-plane hexacopter                       & NO       & NO       & YES          & 0.62\%       & 3.75\%       & 3.12\%       & 2.50\% \\
design\_15317 & hexacopter                                 & NO       & NO       & NO           & 26.88\%      & 8.75\%       & 35.62\%      & 23.75\% \\
design\_16763 & non-axisym non-planar off-plane quadcopter & YES      & YES      & YES          & 0.00\%       & 23.75\%      & 6.25\%       & 10.00\% \\
design\_18368 & hexacopter                                 & NO       & NO       & NO           & 16.25\%      & 0.62\%       & 0.62\%       & 5.83\% \\
design\_18802 & non-axisym non-planar off-plane hexacopter & YES      & YES      & YES          & 1.25\%       & 2.50\%       & 1.25\%       & 1.67\% \\
design\_18914 & non-axisym non-planar off-plane            & YES      & YES      & YES          & 0.62\%       & 0.62\%       & 0.00\%       & 0.42\% \\
design\_18952 & non-axisym non-planar off-plane hexacopter & YES      & YES      & YES          & 0.62\%       & 0.00\%       & 8.12\%       & 2.92\% \\
design\_19051 & non-axisym hexacopter                      & YES      & NO       & NO           & 0.00\%       & 11.25\%      & 3.75\%       & 5.00\% \\
design\_20320 & off-plane hexacopter                       & NO       & NO       & YES          & 0.00\%       & 1.25\%       & 0.00\%       & 0.42\% \\
\textbf{design\_20985*} & off-plane hexacopter                       & NO       & NO       & YES          & 0.62\%       & 41.25\%      & 24.38\%      & 22.08\% \\
design\_24246 & off-plane quadcopter                       & NO       & NO       & YES          & 0.00\%       & 3.75\%       & 3.12\%       & 2.29\% \\
design\_25633 & hexacopter                                 & NO       & NO       & NO           & 1.88\%       & 0.00\%       & 2.50\%       & 1.46\% \\
design\_26003 & non-axisym non-planar off-plane hexacopter & YES      & YES      & YES          & 1.25\%       & 0.00\%       & 0.00\%       & 0.42\% \\
design\_26842 & quadcopter                                 & NO       & NO       & NO           & 33.12\%      & 1.25\%       & 6.25\%       & 13.54\% \\
design\_27369 & non-axisym non-planar off-plane hexacopter & YES      & YES      & YES          & 6.25\%       & 1.25\%       & 4.38\%       & 3.96\% \\
design\_27901 & octocopter                                 & NO       & NO       & NO           & 10.62\%      & 0.00\%       & 0.62\%       & 3.75\% \\
\hline
\caption{Design Features Table - Identifying Features and Pick Percentage. * indicates the Pareto-TOPSIS optimal design. }
\label{tab:drone_feature_3}
\end{longtable}

\end{landscape}

\newpage

\section{Participants' Choice Transition}

\subsection{Study 1}

\begin{figure}[H]
    \centering
    \includegraphics[width=0.5\linewidth]{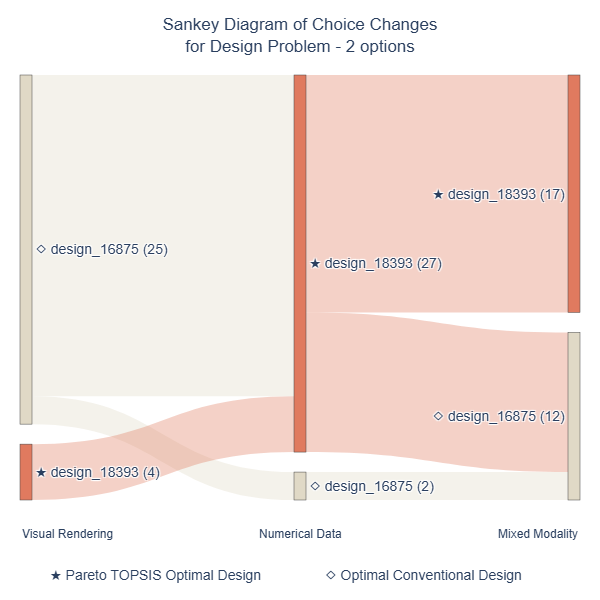}
    \caption{Study 1: Participants' choice transition in the design problem with 2 options. }
    \label{fig:sankey_cmu_flyer_problem_1.png}
\end{figure}

\begin{figure}[H]
    \centering
    \includegraphics[width=0.5\linewidth]{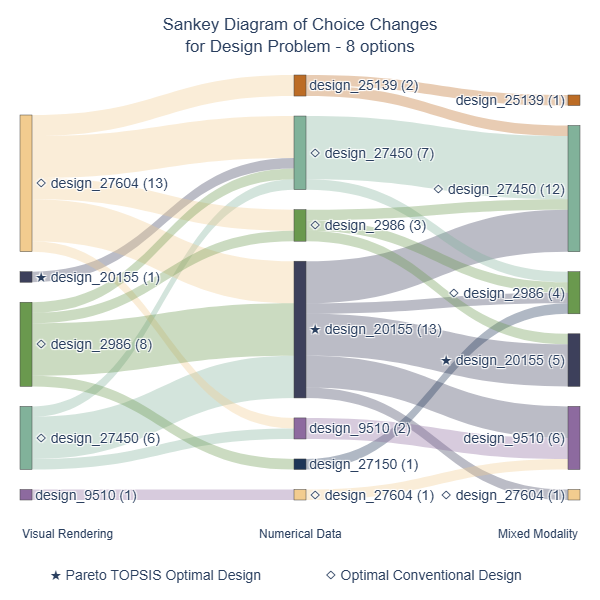}
    \caption{Study 1: Participants' choice transition in the design problem with 8 options. }
    \label{fig:sankey_cmu_flyer_problem_2.png}
\end{figure}

\begin{figure}[H]
    \centering
    \includegraphics[width=0.5\linewidth]{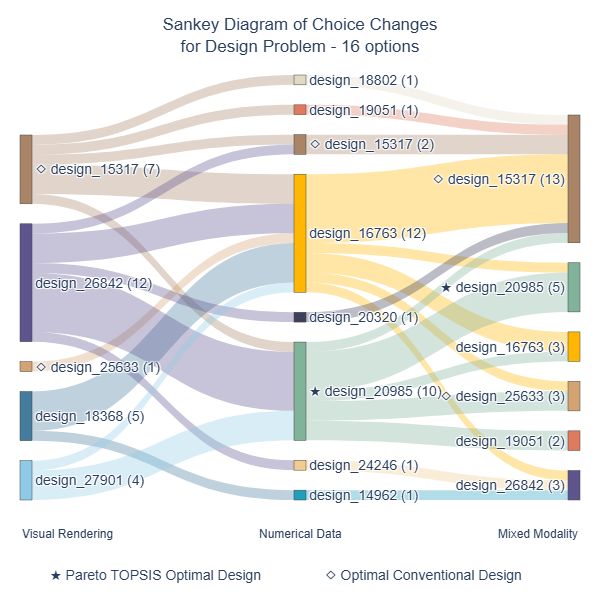}
    \caption{Study 1: Participants' choice transition in the design problem with 16 options. }
    \label{fig:sankey_cmu_flyer_problem_3.png}
\end{figure}

\subsection{Study 2}

\begin{figure}[H]
    \centering
    \includegraphics[width=0.5\linewidth]{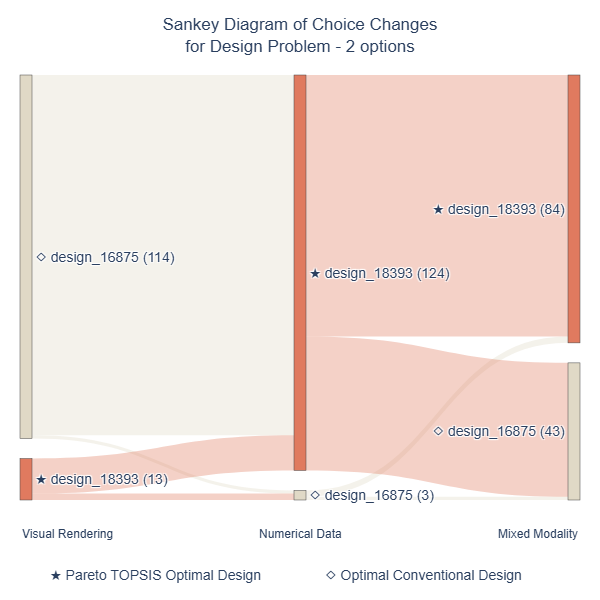}
    \caption{Study 2: Participants' choice transition in the design problem with 2 options. }
    \label{fig:sankey_cmu_class_comb_problem_1.png}
\end{figure}

\begin{figure}[H]
    \centering
    \includegraphics[width=0.5\linewidth]{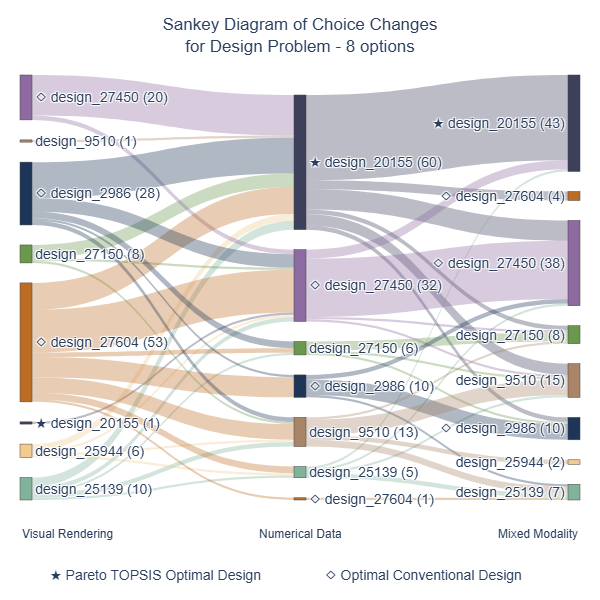}
    \caption{Study 2: Participants' choice transition in the design problem with 8 options. }
    \label{fig:sankey_cmu_class_comb_problem_2.png}
\end{figure}

\begin{figure}[H]
    \centering
    \includegraphics[width=0.5\linewidth]{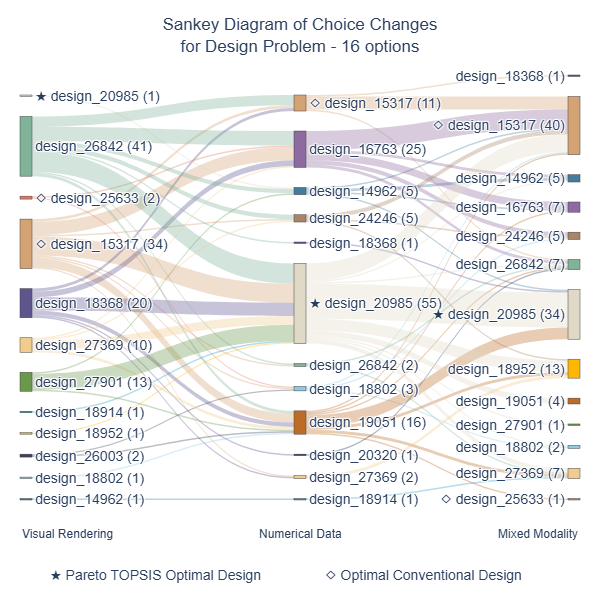}
    \caption{Study 2: Participants' choice transition in the design problem with 16 options. }
    \label{fig:sankey_cmu_class_comb_problem_3.png}
\end{figure}

\newpage

\end{document}